\RequirePackage{ifpdf}
\documentclass[11pt, a4paper]{article}
\usepackage{jheppub}
\usepackage{epsfig}
\usepackage[utf8]{inputenc}
\usepackage{bbold,amsfonts}
\usepackage{graphicx}
\usepackage{amssymb,amsmath}
\usepackage{fancybox,framed}
\usepackage{dsfont}
\usepackage{braket}
\usepackage{slashed}
\usepackage{comment}

\usepackage{soul}

\csname @addtoreset\endcsname{equation}{section}

\def\bseq{\begin{subequation}}  
\def\eseq{\end{subequation}}
\def\bsea{\begin{subeqnarray}}  
\def\esea{\end{subeqnarray}}

\newcommand{\beq}{\begin{equation}}
\newcommand{\bea}{\begin{eqnarray}}
\newcommand{\eea}{\end{eqnarray}}
\newcommand{\eeq}{\end{equation}}

\renewcommand{\a}{\alpha}

\renewcommand{\d}{\delta}
\newcommand{\pa}{\partial}

\newcommand{\G}{\Gamma}

\newcommand{\D}{\Delta}

\newcommand{\m}{\mu}
\newcommand{\M}{\Mu}
\newcommand{\n}{\nu}

\newcommand{\p}{\pi}

\renewcommand{\r}{\rho}

\newcommand{\s}{\sigma}
\renewcommand{\S}{\Sigma}

\renewcommand{\o}{\omega}
\renewcommand{\O}{\Omega}

\newcommand{\mt}[1]{\textrm{\tiny #1}}
\newcommand{\non}{\nonumber \\}
\newcommand{\reef}[1]{(\ref{#1})}
\newcommand{\bt}{\beta}
\newcommand{\eg}{{\it e.g.,}\ }
\newcommand{\ie}{{\it i.e.,}\ }
\def\({\left(} \def\){\right)}
\def\[{\left[} \def\]{\right]}
\def\del{{\partial}}
\def\M{\mathcal{M}}

\newcommand{\see}{S_\textup{EE}}
\newcommand{\tr}{\textup{Tr}}

\newcommand{\al}{\alpha}

\newcommand{\ga}{\gamma}
\newcommand{\de}{\delta}
\newcommand{\ep}{\epsilon}

\renewcommand{\th}{\theta}

\newcommand{\la}{\lambda}

\renewcommand{\r}{\rho}

\renewcommand{\o}{\omega}
\renewcommand{\O}{\Omega}

\newcommand{\ren}{R\'enyi}
\newcommand{\te}{t_\mt{E}}

\declareslashed{}{\backslash}{0}{0}{\Omega}
\declareslashed{}{\backslash}{0}{0}{O}

\author[a,b]{Lorenzo Bianchi}
\author[c,d]{Marco Meineri}
\author[d]{Robert C. Myers}
\author[e]{Michael Smolkin}

\affiliation[a]{Institut f\"ur Physik,
Humboldt-Universit\"at
zu Berlin\\ 
Zum Gro\ss en Windkanal 6, 12489 Berlin, Germany}

\affiliation[b]{II. Institut f\"ur Theoretische Physik,
Universit\"at Hamburg\\
Luruper Chaussee 149,
22761 Hamburg, Germany}

\affiliation[c]{Scuola Normale Superiore, Piazza dei Cavalieri 7 I-56126 Pisa, Italy\\
and Istituto Nazionale di Fisica Nucleare - sezione di Pisa}

\affiliation[d]{Perimeter Institute for Theoretical Physics\\
31 Caroline Street North, ON N2L 2Y5, Canada}

\affiliation[e]{Center for Theoretical Physics and Department of Physics,\\
University of California, Berkeley, CA 94720, U.S.A.
}

\abstract{We propose a field theoretic framework for calculating the dependence of \ren\ entropies on the shape of the entangling surface in a conformal field theory. Our approach rests on regarding the corresponding twist operator as a conformal defect and in particular, we define the displacement operator which implements small local deformations of the entangling surface. 
We identify a simple constraint between the coefficient defining the two-point function of the displacement operator and the conformal weight of the twist operator, which consolidates a number of distinct conjectures on the shape dependence of the \ren\ entropy. As an example, using this approach, we examine a conjecture regarding the universal coefficient associated with a conical singularity in the entangling surface for CFTs in any number of spacetime dimensions. We also provide a general formula for the second order variation of the \ren\ entropy arising from small deformations of a spherical entangling surface, extending Mezei's results for the entanglement entropy.
}

\preprint{DESY 15-229}

\title{\ren\ entropy and conformal defects.}


\begin{document}

\maketitle

\section{Introduction}
\label{sec:intro}

There has been a growing interest in entanglement and \ren\ entropies as probes of complex interacting quantum systems in a variety of areas ranging from condensed matter physics, \eg \cite{wenx,kitaev,haldane} to quantum gravity, \eg \cite{qg1,qg2,qg3,qg4}. While commonly regarded as a useful theoretical diagnostic, the past year has seen remarkable experimental advances where the \ren\ entropy, as well as quantum purity and mutual information, of a system of delocalized interacting particles can be measured in the laboratory \cite{expEE}. This experimental breakthrough strengthens the motivation to develop further theoretical insight into these entanglement measures, particularly in the framework of quantum field theory (QFT). In this paper, we focus our attention on \ren\ entropies \cite{renyi1,renyi2} in the context of conformal field theories (CFTs). Conformal symmetry obviously introduces additional constraints on the \ren\ entropy beyond those in a general QFT, however, CFTs are still a very 
important class of QFTs since they describe physics at the quantum critical points and also shed light on the structure of gravity through the AdS/CFT duality.

In the case of holographic CFTs with a gravitational dual, the Ryu-Takayanagi prescription \cite{Ryu:2006bv,Ryu:2006ef} --- and its generalizations \cite{Hubeny:2007xt,Hung:2011xb,deBoer:2011wk,Dong:2013qoa,Camps:2013zua} --- provides an elegant and practical tool to evaluate the entanglement entropy  across an arbitrary entangling surface.  While the recent derivation \cite{Lewkowycz:2013nqa} of this prescription presents a generalization to holographic \ren\ entropies in principle, explicit holographic calculations of the \ren\ 
entropy have been largely restricted to a spherical entangling surface \cite{Casini:2011kv, Hung:2011nu} --- however, see recent progress in \cite{dong33}. Similarly, efficient computational tools to study the R\'enyi entropies for more general CFTs are rare. Numerical techniques have been developed to evaluate the \ren\ entropy in lattice models describing critical theories, \eg \cite{roger1,roger2,roger3} but these are demanding and must be adapted for the specifics of a given model. Beyond these numerical studies, the existing literature considers primarily the R\'enyi entropy across a spherical entangling surface for a CFT living in flat space, \eg \cite{Casini:2011kv,Casini:2010kt,Perlmutter:2013gua, Hung:2014npa, Lee:2014zaa, Lewkowycz:2014jia}.  

In this paper, we build on conformal defect techniques to develop a field theoretic framework which allows for quantitative studies of the R\'enyi entropy. Conformal defects have a long story, both in two and higher dimensions --- see e.g. \cite{Cardy:1984bb,McAvity:1995zd,billo:confdef}. In section \ref{sec:twistdef}, we begin from the basic definitions to draw a parallel between conformal defects and the twist operators, which enter the calculation of the R\'enyi entropy. This perspective  demonstrates that the R\'enyi entropy readily lends itself to the application of defect CFT techniques. In particular, we define the so-called displacement operator for the twist operators, which then implements small deformations of the entangling surface.

Hence the displacement operator can be used for perturbative calculations of the \ren\ entropy when small modifications are made in the geometry of the entangling surface. 
Our focus on the displacement operator in the following arises because recently there has been a great deal of interest in the shape dependence of R\'enyi  and entanglement entropies,  \eg \cite{mark0,mark1,Bueno:2015rda,Bueno:2015xda,Bueno:2015qya,Bueno:2015lza,safdi1,aitor7}.
First, we show below for a planar or spherical entangling surface that the second order variation of the \ren\  entropy is fixed by the two-point function of the displacement operator. Given this framework, one of our key results is then to identify a simple relation between the coefficient defining the two-point function of the displacement operator and the conformal weight of the twist operator, which unifies a number of distinct conjectures with regards to the shape dependence of the \ren\ entropy. 
In particular, in section \ref{sec:defsphere}, we apply this approach to evaluate the second order variation of the \ren\  entropy for a spherical region. In the limit that the \ren\ index goes to one, using the new relation, we precisely recover Mezei's conjecture for variations of the entanglement entropy \cite{mark0} --- see also \cite{mark1}. Further the displacement operator can also be used to examine the variation of the \ren\ entropy for small but `singular' deformations of the entangling surface. In section \ref{sec:cusp}, we consider the case of a planar entangling surface which undergoes a singular deformation to create a small conical singularity. With the previous relation, our result for the change in the R\'enyi entropy matches previous conjectures with regards to cusp and cone geometries in the limit that the entangling surface is almost smooth \cite{Bueno:2015rda,Bueno:2015xda,Bueno:2015qya,Bueno:2015lza}.
In section \ref{sec:fourd}, we focus on the R\'enyi entropy across an arbitrary entangling surface in four spacetime dimensions.  We are able to relate two coefficients in the universal contribution to the \ren\ entropy to the conformal weight  of the twist operator and to the coefficient in the two-point function of the displacement operator, respectively. Our relation between the latter two quantities then yields the equality of these coefficients, as was conjectured for all four-dimensional CFTs in \cite{safdi1}. Hence, interestingly, the relation between the coefficient of the two-point function of the displacement and the conformal weight of the twist operator underlies a number of existing conjectures in the literature about the \ren\  entropy. 

However, at this point, we must add that recent holographic calculations \cite{dong34} imply that the proposed relation does {\it not} hold for general values of the \ren\ index in four-dimensional holographic CFTs dual to Einstein gravity. Hence it becomes an interesting question to ask for precisely which CFTs does such a constraint hold. For example, our calculations in Appendix \ref{sec:scalar}  confirm that it does in fact hold for free massless scalars in four dimensions. 

In section \ref{sec:singularity}, we pose the question whether it may be possible to define the twist operator through the operator product expansion generically available in the presence of defects, and we point out an intriguing universal 
feature of the fusion of the stress-tensor with this specific extended operator. 
We conclude with a discussion of our results and possible future directions in section \ref{sec:discussion}. We also have a number appendices where we present various technical calculations whose results are used in the main text. In Appendix \ref{sec:ward}, we derive a set of Ward identities in the presence of a twist operator. In Appendix \ref{varmet}, we list some useful formulas which describe variation of various geometrical objects around the flat space. We devote Appendix \ref{sec:scalar} to the example of a four-dimensional free scalar field, where the displacement operator can be given a precise identity in terms of the elementary field.

Finally, let us add that while this paper was in the final stages of preparation, ref.~\cite{new} appeared. Although their discussion only considers the entanglement entropy, some of the results overlap with aspects of the present paper. 

\section{Twist operators as conformal defects}
\label{sec:twistdef}

A central object for our discussion will be the twist operator which naturally arises in evaluating {R\'enyi entropies} in quantum field theory \cite{Hung:2014npa,Calabrese:2004eu,Cardy:2007mb}.\footnote{In two-dimensional CFTs, twist fields associated to branch points were first introduced in~\cite{Knizhnik:1987xp,Dixon:1986qv}.} Therefore, let us start by recalling the definition of our main player. We begin with a generic QFT in flat $d$-dimensional spacetime. On a given time slice, the QFT is in a global state described by the density matrix $\r$ --- in fact, shortly we will restrict our attention to the vacuum state. We consider the density matrix $\r_A$ obtained when the state is restricted to a particular region $A$, \ie obtained by tracing over the degrees of freedom in the complementary region  $\bar A$ of the time slice:
\begin{equation}
\r_A = \tr_{\bar A}(\r)\,. \label{reduce}
\end{equation}
The one-parameter family of R\'enyi entropies associated to the reduced density matrix $\r_A$ is defined as follows \cite{renyi1,renyi2}:
\begin{equation}
S_n = \frac{1}{1-n}\, \log \tr (\r_A^n).
\label{renyi}
\end{equation}
The entanglement entropy, \ie the von Neumann entropy, is recovered with the limit:
\begin{equation}
\lim_{n\to 1} S_n =\see = - \tr (\r_A \log \r_A)\, . \label{sea}
\end{equation}
Here we have implicitly considered the \ren\ index $n$ in \reef{renyi} to be a real number. However, specifically for integer $n$ (with $n>1$), a path integral construction, which is widely known as the replica trick, allows us to evaluate the R\'enyi entropies for a QFT. An analytic continuation is then required to make contact with the entanglement entropy but we will have nothing to add about the conditions under which the continuation is reliable.

The replica trick begins by evaluating the reduced density matrix $\r_A$ in terms of a (Euclidean) path integral on $\mathbb{R}^d$ but with independent boundary  conditions fixed over the region $A$ as it is approached from above and below in Euclidean time, \eg with $\te\to0^\pm$. The expression $\tr (\r_A^n)$ is then evaluated by extending the above to a path integral on a $n$-sheeted geometry \cite{Calabrese:2004eu,Cardy:2007mb}, where the consecutive sheets are sewn together on cuts running over $A$. Denoting the corresponding partition function as $Z_n$,  we can write the \ren\ entropy \reef{renyi} as
\begin{equation}
S_n = \frac{1}{1-n}\ \log \frac{Z_n}{Z_1^n}\,,
\label{sn}
\end{equation}
where the denominator $Z_1^n$ is introduced here to ensure the correct normalization, \ie $\tr[\rho_A]=1$.
The partition function $Z_n$ has an important symmetry. That is, even if in the above construction we chose to glue the copies together along the codimension-one submanifold $A$ on the $\te=0$ slice, the precise location of the cut between different sheets is meaningless --- see for instance section 3.1 of \cite{Caraglio:2008pk}. Hence the only source of breaking of translational invariance on each sheet is at the location of the entangling surface, \ie the boundary of $A$. Since the modification is local in this sense, it can be reinterpreted as the insertion of a \emph{twist operator} $\tau_n$. In defining $\tau_n$, the above construction is replaced by a path integral over $n$ copies of the underlying QFT on a single copy of the flat space geometry.  The twist operator is then defined as a codimension-two surface operator in this $n$-fold replicated QFT, which extends over the entangling surface and whose expectation value yields 
\begin{equation}
\braket{\tau_n}\equiv \frac{Z_n}{Z_1^n} = e^{(1-n) S_n}.
\label{sigman}
\end{equation}
Hence eq.~\reef{sigman} implies that $\tau_n$ opens a branch cut over the region $A$ which then connects consecutive copies of the QFT in the $n$-fold replicated theory. Note that here and in the following,  we omit the $A$ dependence of $\tau_n$ to alleviate the notation. 

In proceeding, we restrict our attention to the case where the QFT of interest is a conformal field theory and the state is simply the flat space vacuum state. Now let us take a closer look at the residual symmetry group in the presence of the twist operator. In doing so, we restrict ourselves to a very symmetric situation where we choose $A$ to be half of the space. That is, we choose $\tau_n$ to lie on a flat ($d\!-\!2$)-dimensional plane, which we denote as $\Sigma$.  For concreteness, we parametrize  $\mathbb{R}^d$ with coordinates $(x^1,\dots, x^d)$, and we locate the twist operator at $x^1=0=x^2$. In the following, we will denote directions orthogonal to $\Sigma$ with Latin indices from the beginning of the alphabet ($a,\,b,\dots$) and parallel directions with Latin indices from the middle of the alphabet ($i,\ j, \dots$), while $\m=(i,a)$. Let us explicitly notice that since a spherical entangling surface can be obtained from the planar one by means of a conformal transformation, the following applies 
equally well to the spherical case. 

Now, the stabilizer of a $(d\!-\!2)$-dimensional  plane within the $d$-dimensional conformal algebra is the subalgebra $so(d-1,1)\times u(1)$. The first factor comprises  the conformal transformations in $(d\!-\!2)$ dimensions, while the second consists of rotations in the transverse space. Let us choose the cut to lie along a half-plane in $\mathbb{R}^d$, \eg $x^{1}<0$ (and $x^2=0$), then a moment's thought is sufficient to realize that the gluing condition is preserved only if the same conformal transformation is applied to all the copies at the same time.\footnote{In the density matrix language, this is rephrased in the statement that the transformation $U\r_A U^{-1}$ is a symmetry of $Z_n$ only if applied to the $n$ factors of $\r_A$ appearing in the trace \eqref{renyi}.} The rotations in the transverse plane, on the other hand, move the cut, which can be brought back to the original position through the symmetry of the partition function which we referred to above. This leads to a remark on the 
structure of 
the symmetry group. A rotation of an angle $2\pi$ has the net effect of shifting by one the labeling of the replicas: in a correlation function, an operator inserted in the $i$-th copy 
ends up in the $(i+1)$-th one. Therefore, the $u(1)$ algebra exponentiates in the $n$-fold cover of the group $O(2)$. 
Up to this subtlety, we see that the symmetry group preserved by the twist operator is the same as the one preserved by a flat conformally invariant extended operator, \ie a conformal defect. 

The symmetry algebra places constraints on observables, which in many cases have been worked out in the context of defect CFTs. In particular, this is the case for correlation functions of local operators \cite{billo:confdef}. The twist operator is a conformal defect placed in the tensor product $(\textup{CFT})^n$ rather than in the original conformal field theory. Therefore, it is especially interesting to consider the consequences of interactions among replicas, which distinguish this setup from a mere local modification of a CFT on $\mathbb{R}^d$: these are probed by correlation functions of operators belonging to different copies of the theory. Such correlators
do not escape the defect CFT framework and in particular can be handled with the classical tool available in any conformal field theory: the existence of an operator product expansion (OPE), which converges inside correlation functions. In the presence of a defect, bulk excitations can be brought close to the extended operator, and be expressed as a sum over local operators on the defect. This corresponds to a new OPE channel, usually referred to as the 
defect OPE. If we denote defect operators with a hat, the defect OPE of a bulk scalar of scaling dimension $\D$ takes the following form:
\begin{equation}
O(x^a,x^i) \sim b_{\widehat{O}_0}\, r^{\widehat{\D}_0-\D} \widehat{O}_0(x^i) +\dots, \qquad r\equiv |x^a|.
\label{defopeloose}
\end{equation}
The meaning of the label $0$ given to the defect operator will become clear in a moment. Let us stress that operators in our formulae will always be thought of as inserted on a single copy of the CFT and when present, sums over replicas will be written explicitly. Now consider a correlator of bulk primaries which belong to different factors of the $n$-fold replicated CFT. We can substitute to each of them the respective defect OPE, and since the latter converges inside correlation functions, the resulting sum over two-point functions of defect operators must reproduce the original correlator. In particular, we see that the expression on the right hand side of eq. \eqref{defopeloose} must retain the information about the copy in which the primary on the left hand side was inserted. This is possible thanks to the global structure of the symmetry group --- that is, the fact that rotations around the defect are combined non-trivially with the $\mathbb{Z}_n$ replica symmetry. The rotational symmetry around an 
extended 
operator is a global symmetry from the point of view of the theory on the defect. As a consequence, defect operators carry a $u(1)$ quantum number $s$. In our case, this transverse spin is rational: $s=k/n$, $k$ being an integer. We see that the defect OPE of a bulk scalar contains in general terms of the form
\begin{equation}
O(x^a,x^i) \sim b_{\widehat{O}_s}\, |x^a|^{\widehat{\D}_s-\D} e^{i s \phi} \widehat{O}_s(x^i) +\dots
\label{defopespin}
\end{equation}
where $\phi \in [0,2\pi n)$ is the angle in a plane orthogonal to the defect and provides the information about the replica on which the bulk primary has been inserted. In appendix \ref{sec:scalar}, we shall see explicit examples of OPEs of the form \eqref{defopespin} in the free scalar theory, and how they allow us to decompose correlation functions of bulk primaries placed in arbitrary positions.

The breaking of translational invariance in the directions transverse to the entangling surface gives rise to an operator of transverse spin $s=1$, which is always present on defects in local theories. Indeed, the Noether current which generates translations fails to be conserved only at the position of the defect, so that a new contact term should be present in the 
Ward identities of the stress-tensor. This defines the displacement operator $D^a$:
\begin{equation}
\sum_{{\rm m}=1}^n\pa_\m T_\mt{(m)}^{\m a}(x^\nu) = \d_\Sigma(x)\ D^a(x^i),
\label{stressward}
\end{equation}
where the index m runs over the replicas\footnote{We stress again that in general, our calculations will refer to bulk operators in a single copy of the replicated CFT. Hence $T_\mt{(m)}^{\m \nu }$ here denotes the stress tensor in the m'th copy of the CFT and the total stress tensor for the full theory would be given by $T_\mt{tot}^{\m \nu }=\sum_{{\rm m}=1}^nT_\mt{(m)}^{\m \nu }$. However, in order to reduce  clutter in expressions below, we will drop the subscript (m) but $T^{\m \nu }$ still denotes the single-copy stress tensor. The total stress tensor will always be denoted as $T_\mt{tot}^{\m \nu }$.} and $\de_\Sigma$ denotes the delta function in the transverse space with support on the  $\Sigma$. The sum over replicas appears because, as mentioned, symmetry transformations should be applied to all the sheets in the same way, resulting in a sum over insertions of the stress-tensor.
Eq. \eqref{stressward} is written in a somewhat loose notation, which highlights the properties of the displacement. The right hand side should be intended as an additional contribution arising when the left hand side is inserted in a correlation function. We refer to appendix \ref{sec:ward} for a derivation, and we content ourselves here with a few remarks: It is important that the quantum numbers of $D^a$ are fixed by the Ward identity, \ie its scaling dimension is $\D=d-1$ and it carries one unit of spin under rotations around the defect. Notice that its normalization is also fixed by \eqref{stressward}. Therefore, its Zamolodchikov norm $C_D(n)$ is a property of the defect under consideration:
\begin{equation}\label{2ptdisp}
\braket{D^a(x^i)\,D^b(0)}_n = C_D(n)\ \frac{\d^{ab}}{|x^i|^{2(d-1)}}\,.
\end{equation}
Here and in the following the subscript $n$ applied to expectation values implies the presence of the twist operator:
\begin{equation}
\braket{O}_n \equiv \frac{\braket{\tau_n\, O}}{\braket{\tau_n}}.
\end{equation}
Let us finally mention that generic defects might have a more complicated structure of contact terms showing up in the divergence of the stress-tensor: more operators might be present, associated with derivatives of $\de$-functions appearing in eq. \eqref{stressward}. They can be written down systematically  \cite{billo:confdef} but we will not need this information here.

A consequence of the Ward identity \eqref{stressward} is that a small deformation $\d x^a (x^i)$ of the defect, is obtained by integrating the displacement operator in the action. The first order variation under such a deformation can be written
\begin{equation}
\d \braket{X}_n = - \int\! d^{d-2}x\, \d x^a(x^i) \braket{D^a(x^i) X}_n.\label{211}
\end{equation}
where $X$ is an arbitrary product of local operators. As already pointed out, a flat twist operator preserves a subgroup of the conformal transformations which includes dilatations. As an immediate consequence, scale invariance prevents defect operators from acquiring an expectation value in this particular case. Hence, the first order variation of the partition function \eqref{sigman} vanishes for a flat (or spherical) entangling surface, or more precisely, it is non-universal.
The second order variation is then related directly to $C_D$. Indeed, denoting  the variation as $\ep\, \d x^a$, we find
\begin{equation}\label{deformation}
\left. \frac{1}{\braket{\tau_n}}\frac{d^2}{d \ep^2} \braket{\tau_n} \right|_{\ep=0}
= \int\! d^{d-2}x \int\! d^{d-2}x' \braket{D^a(x) D^b(x')}_n \d x_a \d x'_b.
\end{equation}
The double integration will contain divergences which must be regulated. However, power-law divergences can be unambiguously tuned away, and finite or logarithmically divergent parts are universal well defined quantities, proportional to $C_D$. 

Hence eq.~\reef{deformation} shows very explicitly that the displacement operator is the key element of the defect CFT living on the twist operator, which governs  the shape dependence of the R\'enyi entropy, which has been extensively studied in the recent literature \eg \cite{mark0}--\cite{aitor7}. 
A key result of this paper is unify a variety of conjectures related to this shape dependence in terms of a constraint on $C_D$, the coefficient defining the two-point function \reef{2ptdisp} of the displacement operator. In particular, these conjectures imply that the value of $C_D$ is entirely determined by the one-point function of the stress tensor in presence of the defect, also called the conformal dimension of the twist operator. The latter, dubbed $h_n$, is defined by the leading singularity of the one-point function $\braket{T_{\m\n}}_n\equiv\braket{T_{\m\n}\,\tau_n}/\braket{\tau_n}$. For a planar conformal defect in Euclidean flat geometry, this leading singularity is easily identified as it is completely fixed by symmetry 
\begin{align}\label{eq:oneptT}
 \braket{T_{ij}}_n&=-\frac{h_n}{2\p n}\frac{\d_{ij}}{r^d}\, ,  & \braket{T_{aj}}_n&=0\,, &  \braket{T_{ab}}_n&=\frac{h_n}{2\p n}\frac{(d-1)\, \d_{ab}-d\, n_an_b}{r^d}\, .
\end{align}
Here $n_a$ is a unit normalized vector normal to the entangling surface and $r=|x^a|$ the transverse distance. The factor $n$ in the denominator appears so that $h_n$ is the coefficient in the one-point function for the total stress tensor (summed over all of the replicas), \eg as defined in   \cite{Hung:2014npa}. In the following, we demonstrate that if, in a $d$-dimensional CFT, the values of $C_D(n)$ and $h_n$ are constrained to obey the following equality
\begin{equation}\label{eq:CDhrel}
 C_D(n)= d\, \Gamma\!\left(\tfrac{d+1}{2}\right)\, \left(\tfrac{2}{\sqrt{\pi}}\right)^{d-1}\, h_n, 
\end{equation}
then the \ren\ entropy satisfies a number of interesting properties, outlined below, with regards to shape dependence.

One immediate consequence of this relation is $C_D(1)=0$, which must hold since the defect disappears for $n=1$. Further, if we analytically continue  \eqref{eq:CDhrel} to real $n$, we can consider the first order variation around $n=1$:
\begin{align}\label{CDhrelder}
 \pa_n C_D|_{n=1}&=d\, \Gamma\!\left(\tfrac{d+1}{2}\right)\, \left(\tfrac{2}{\sqrt{\pi}}\right)^{d-1}\, \pa_n h_n|_{n=1}=\frac{2\pi^2}{d+1} C_T,
\end{align}
where we used the relation
\begin{equation}\label{hCTrel}
 \pa_n h|_{n=1}=2\pi^{\frac{d}{2}+1}\frac{\Gamma(\frac{d}{2})}{\Gamma(d+2)}C_T,
\end{equation}
first observed in~\cite{Hung:2011nu} for holographic theories and then proven in~\cite{Hung:2014npa} for general CFTs. Implicity, the recent results of \cite{new} imply that eq.~\reef{CDhrelder} holds for generic CFTs. Hence in the vicinity of $n=1$, the proposed relation \reef{eq:CDhrel} is a constraint that holds for general CFTs.

Moving away from $n=1$, the constraint in eq.~\reef{eq:CDhrel} produces an number of interesting properties for the shape dependence of the \ren\ entropy, which have appeared previously in the literature as conjectures: 
\begin{itemize}
 \item In section \ref{sec:defsphere}, we calculate the second order correction to the R\'enyi entropy induced by small perturbations of a perfect sphere. In the limit $n\to 1$, the formula \eqref{CDhrelder} reproduces the variation of the entanglement entropy across a deformed sphere conjectured in~\cite{mark0} for arbitrary dimensions, which was recently proven in \cite{new}. 
 \item Eq.~\eqref{CDhrelder} also allows one to compute the universal contribution to the R\'enyi entropy for an entangling surface with a (hyper)conical singularity of opening angle $\O$. The leading coefficient in an expansion around the smooth entangling surface has been conjectured to be related the conformal weight $h_n$~\cite{Bueno:2015lza} --- see also \cite{Bueno:2015rda,Bueno:2015xda,Bueno:2015qya}. In section \ref{sec:cusp}, we prove the equivalence of that conjecture and formula~\eqref{CDhrelder}.
  \item With $d=4$, eq.~\eqref{eq:CDhrel} implies the equivalence of the coefficients $f_b(n)$ and $f_c(n)$ in the universal part of the four-dimensional R\'enyi entropy for general $n$, as discussed in \cite{Lewkowycz:2014jia,safdi1}. This is demonstrated in section \ref{sec:fourd} by relating $f_b$ to $C_D$, and $f_c$ to $h_n$. However, we re-iterate that \cite{dong34} recently showed that the proposed equivalence $f_b(n)=f_c(n)$ does not hold for four-dimensional holographic CFTs dual to Einstein gravity. 
\end{itemize}
The latter result demonstrates that eq.~\eqref{eq:CDhrel} is {\it not} a universal relation that holds in all CFTs (for general values of $n$). However, it is then interesting to ask for precisely which CFTs does such a constraint hold. It seems that free field theories are a good candidate for such a theory. Certainly, the results of \cite{Bueno:2015qya,Dowker1,Dowker2} imply that eq.~\eqref{eq:CDhrel} holds for free scalars and fermions in three dimensions. Further, our calculations in Appendix \ref{sec:scalar}  confirm that it also holds for free massles scalars in four dimensions. We hope to return to this question in future work \cite{future}. 

\section{R\'enyi and entanglement entropy across a deformed sphere }
\label{sec:defsphere}

In this section, we study shape dependence of the R\'enyi entropy for a generic CFT in flat space. In particular, we calculate the second order correction to the R\'enyi entropy induced by small perturbations of a perfect sphere. In the limit $n\to 1$, our findings agree with the holographic results previously found in \cite{mark0}.

Starting from \reef{deformation}, we note that upon slightly deforming a spherical entangling surface with $\ep\,\delta x^a$, the leading correction to $S_n$ appears at second order and is given by 
\beq 
  \delta S_n=\frac{\ep^2}{2 (1-n)}\int_\Sigma \int_{\Sigma'}\  \langle D^{a}(x) D^b(x')\rangle_n\, \delta x_a \, \delta x'_b + \mathcal{O}(\delta x^3)~.
  \label{Renyi}
\eeq
Here, the two integrals run over the original spherical entangling surface of radius $R$. We will restrict the deformation to the $\te=0$ time slice and denote $\delta  \vec x=  f(x)\, \hat r$ where $\hat r$ is a unit vector in the radial direction. The relevant correlator \reef{2ptdisp} then beomes
\beq
 \langle D^r (x)\, D^r (x')\rangle_n =   {C_D \over (x-x')^{2(d-1)}}={C_D\over (2R^2)^{d-1}}\, {1\over (1-\cos\gamma)^{d-1}}~,
 \label{2p-disp}
\eeq
with $\gamma$ being the angle between $x,x'\in S^{d-2}$. 

Let us now represent the two-point correlator \reef{2p-disp} in the basis of spherical harmonics on $S^N$ ($N\equiv d-2$) 
\beq
 Y_{\ell_N\ldots\ell_1}(\theta_N\ldots\theta_1)={1\over\sqrt{2\pi}} \,e^{i\ell_1\theta_1}\,\prod_{n=2}^N \ _n c_{\ell_n}^{\ell_{n-1}}\, (\sin\theta_n)^{2-n\over 2} \,
 P_{\ell_n+{n-2\over 2}}^{-\({\ell_{n-1}+{n-2\over 2}}\)}(\cos\theta_n)
 \label{harm}
\eeq
 where $\ell_N\geq\ell_{N-1}\geq\ldots\geq |\ell_1|$ are integers and
 \bea
  ds^2_N&=&d\theta_N^2+\sin^2\theta_N ds^2_{N-1}~, \quad ds_1^2=d\theta_1^2 ~,
  \\[1em]
  \sqrt{g}&=&\sin^{N-1}\theta_N\sin^{N-2}\theta_{N-1}\cdots\sin\theta_2~,
  \\[1em]
 P^{-\mu}_{\nu}(x)&=&{1 \over \Gamma(1+\mu)}\( {1-x \over 1+x} \)^{\mu/2}
 \ _{2}F_1\(-\nu ~,~ \nu+1 ~;~1+\mu ~;~ {1-x\over 2}\)
 \quad ,
  \\[1em]
 _n c_{L}^{l}&=&\[ {2L+n-1 \over 2}{(L+l+n-2)! \over (L-l)!} \]^{1/2}
 \quad .
 \label{defs}
 \eea

For simplicity, we assume that one of the points is sitting at the north pole, in which case only spherical harmonics with $\ell_{N-1}=\ell_{N-2}=\ldots=\ell_1=0$ contribute 
\beq
 Y_{\ell_N 0 \ldots 0}(\theta_N)=\sqrt{\Gamma\(N\over 2\) \over 2\pi^{N\over2}}   \ _N c_{\ell_N}^{0}\, (\sin\theta_N)^{2-N\over 2} \,
 P_{\ell_N+{N-2\over 2}}^{-{N-2\over 2}}(\cos\theta_N) ~.
\eeq
Hence, by assumption $\gamma=\theta_N$ in \reef{2p-disp}, and the following identity holds
\bea
  \langle D (x) D (x')\rangle_n &=& {C_D\over (2R^2)^{d-1}} \sum_{\ell_N} A_{\ell_N} Y_{\ell_N 0 \ldots 0}(\gamma) ~,
  \non
  A_{\ell_N}&=& \sqrt{2\pi^{N\over2} \over \Gamma\(N\over 2\)}  \ _N c_{\ell_N}^{0} \int_{-1}^1 dz \,
  {(1-z^2)^{N-2\over 4} \over (1-z)^{N+1}} \, P_{\ell_N+{N-2\over 2}}^{-{N-2\over 2}}(z) 
  \label{Acoeff}
\eea
where we introduced a new variable $z=\cos\gamma$. 

The above integral diverges at $z=1$. This is not surprising given that the coefficients $A_{\ell_N}$ correspond to a spherical harmonic representation of a singular function \reef{2p-disp}. To regulate these coefficients let us modify  the power of $(1-\cos\gamma)$ in \reef{2p-disp} by introducing a new parameter $\al$ such that $A_{\ell_N}$ takes the form
\beq
 A_{\ell_N}={\pi^{N\over 4} \, \ _N c_{\ell_N}^{0} \over 2^{N+1\over 2} \, \Gamma^{3\over 2}\({N\over 2}\)} 
 \lim_{\al\to 0}\int_0^1 dy \, y^{\al-{N+4\over 2}}  \ _2 F_1\Big(-\ell_N-{N\over 2}+1 ~ , ~ \ell_N+{N\over 2} ~ ;~ {N\over 2} ~;~y\Big) ~.
\eeq
where $y=(1-z)/2$ and we used \reef{defs} to express the associated Legendre polynomial in terms of the hypergeometric function. Now the integral can be readily evaluated assuming that $\al$ is large enough\footnote{As usual, small values of $\al$ are treated by analytic continuation.}
\beq
 A_{\ell_N}={\pi^{N\over 4}  \ _N c_{\ell_N}^{0} \over 2^{N+1\over 2} \, \Gamma^{3\over 2}\({N\over 2}\)} 
\lim_{\al\to 0}{ \ _3 F_2 \Big(\al-{N\over 2}-1 ~ , ~ -\ell_N-{N\over 2}+1 ~,~ \ell_N+{N\over 2} ~;~ \al-{N\over 2} ~, ~{N\over 2} ~;~1\Big) \over \al-{N\over 2}-1 }
\label{Acoeff2}
\eeq
For odd $N$ (odd $d$) the limit $\al\to 0$ is finite. However, it diverges for even $N$ (even $d$). Therefore we analyze these cases separately.

\subsection*{Odd $d$}

For odd $d$, we have
\bea
 A_{\ell_N}&=&-{\pi^{N\over 4}  \ _N c_{\ell_N}^{0} \over 2^{N-1\over 2} \, \Gamma^{3\over 2}\({N\over 2}\)} { \ _3 F_2 \Big(-{N\over 2}-1 ~ , ~ -\ell_N-{N\over 2}+1 ~,~ \ell_N+{N\over 2} ~;~ -{N\over 2} ~, ~{N\over 2} ~;~1\Big) \over N+2 }
\non
&=&(-1)^{N-1\over 2} {\pi^{N+4\over 4}  \ _N c_{\ell_N}^{0} \over 2^{N-3\over 2} \, N(N+2)\, \Gamma^{3\over 2}\({N\over 2}\) \Gamma(N+1)} \,  \prod_{k=1,\ldots,N+2}(\ell_N+k-2) 
\eea
Using now the addition theorem for spherical harmonics
\beq
 Y_{\ell_N 0 \ldots 0}(\gamma)= {1\over  _N c_{\ell_N}^{0}}\, \sqrt{(4\pi)^{N\over2} \Gamma\(N\over 2\) \over 2 }  \sum_{\ell_{N-1},\ldots,\ell_1} 
 Y^*_{\ell_N\ldots\ell_1}(x)Y_{\ell_N\ldots\ell_1}(x') ~,
 \label{thm}
\eeq
we obtain from \reef{Acoeff}
\bea
  \langle D (x) D (x')\rangle &=& \, C_D \, { (-1)^{N-1\over 2} \,\pi^{N+2\over 2}  \over 2(2R^2)^{N+1}\Gamma(N+1)\Gamma\({N\over 2} +2 \)} 
    \non
  &\times& \sum_{\ell_N,\ldots,\ell_1}
 Y^*_{\ell_N\ldots\ell_1}(x)Y_{\ell_N\ldots\ell_1}(x')\prod_{k=1,\ldots,N+2}(\ell_N+k-2)   ~.
\eea
Substituting this result into \reef{Renyi}, yields
\beq
 \delta S_n=\epsilon^2 \, {C_D\over (n-1)} \,
   { (-1)^{d-1\over 2} \pi^{d\over 2}  \over 2^{d+1}\,\Gamma(d-1)\Gamma\({d\over 2} +1 \)} 
   \sum_{\ell_N,\ldots,\ell_1}
 |a_{\ell_N\ldots\ell_1}|^2 \prod_{k=1,\ldots,d}(\ell_N+k-2)  + \mathcal{O}(\epsilon^3) \, ~,
 \label{oddSn}
\eeq
where $a_{\ell_N\ldots\ell_1}$ are the coefficients of $f(x)$ in a spherical harmonics representation. This result agrees with \cite{mark0} for any odd $d$ provided that \reef{CDhrelder} holds.

\subsection*{Even $d$}

For even $d$, the limit $\al\to 0$ in \reef{Acoeff2} is singular due to logarithmic divergence. To extract the numerical coefficient of this divergence, we expand the integrand in \reef{Acoeff} around $z=1$ and keep only the logarithmically divergent term:
\beq
 {(1-z^2)^{N-2\over 4} \over (1-z)^{N+1}} \, P_{\ell_N+{N-2\over 2}}^{-{N-2\over 2}}(z) =  
 \frac{ (-1)^{N\over 2}\prod_{k=1,\ldots,N+2}(\ell_N+k-2) }{ 2^{N+2\over 2} \Gamma(N+1)\Gamma\({N\over 2} +2 \)} \, {1\over z-1} + \ldots \, ,
\eeq
The ellipsis denotes terms which do not generate logarithms upon integration.  
Hence,
\beq
 A_{\ell_N}= (-1)^{N+2\over 2}\sqrt{2\pi^{N\over2} \over \Gamma\(N\over 2\)}  \ _N c_{\ell_N}^{0} \,
 \frac{ \prod_{k=1,\ldots,N+2}(\ell_N+k-2) }{ 2^{N\over 2} \Gamma(N+1)\Gamma\({N\over 2} +2 \)} \log(R/\delta) + \ldots \, ,
\eeq
with $\delta=R\cdot\delta\gamma$ being the short-distance cut-off. Using now \reef{thm}, we obtain 
\bea
  \langle D (x) D (x')\rangle_n &=&  \log(R/\delta) \, C_D \, { (-1)^{N+2\over 2} \,\pi^{N\over 2}  \over (2R^2)^{N+1}\Gamma(N+1)\Gamma\({N\over 2} +2 \)} 
    \non
  &\times& \sum_{\ell_N,\ldots,\ell_1}
 Y^*_{\ell_N\ldots\ell_1}(x)Y_{\ell_N\ldots\ell_1}(x')\prod_{k=1,\ldots,N+2}(\ell_N+k-2)  + \ldots \, ~.
\eea
Substituting this result into \reef{Renyi}, yields
\bea
 \delta S_n&=&\epsilon^2 \, {C_D\over (n-1)} \,
   { (-\pi)^{d-2\over 2}  \over 2^{d}\,\Gamma(d-1)\Gamma\({d\over 2} +1 \)} \,\log(R/\delta)
    \non
  && \qquad\times\ \sum_{\ell_N,\ldots,\ell_1}
 |a_{\ell_N\ldots\ell_1}|^2 \prod_{k=1,\ldots,d}(\ell_N+k-2)  + \ldots \, ~,
\eea
where $a_{\ell_N\ldots\ell_1}$ are coefficients of $f(x)$ in a spherical harmonics representation. Combined with \reef{CDhrelder}, this result is again in full agreement with \cite{mark0}.

\section{The cone conjecture}\label{sec:cusp} 

In this section, we consider the relation of the proposed constraint \reef{eq:CDhrel} to various conjectures about the universal contribution to the \ren\  entropy coming from singular deformations of entangling surfaces. In particular, \cite{Bueno:2015rda,Bueno:2015xda} proposed a conjecture for the universal corner contribution to the  entanglement entropy in three-dimensional CFTs, and this conjecture was then extended to R\'enyi entropy in \cite{Bueno:2015qya}. Finally, the discussion was extended to higher dimensions in \cite{Bueno:2015lza}. In order to introduce the claim of these conjectures, we consider a deformation of a flat entangling surface which consists in creating a conical singularity. The three- and four-dimensional cases are shown in figure 1 of ref.~\cite{Bueno:2015lza}. The universal contribution to the R\'enyi (and consequently the entanglement) entropy is affected by this modification. In particular, if the twist operator is smooth, the universal contribution would be logarithmically divergent in even dimensions and constant (\ie regulator independent) in odd dimensions. When a conical singularity is present an additional 
logarithm emerges and the universal contribution to the R\'enyi entropy takes the form
\begin{equation}
 S_n^{\text{univ}}(A)=\left\{\begin{array}{ll}
                          (-1)^{\frac{d-1}{2}}\ a_n^{(d)}(\Omega)\ \log (\ell/\d) &\qquad d \text{  odd}\\
                          (-1)^{\frac{d-2}{2}}\ a_n^{(d)}(\Omega)\ \log^{2\,}\! (\ell/\d)  &\qquad  d \text{  even}
                          \end{array}
\right. \label{wawa}
\end{equation}
Here $\Omega$ is the opening angle of the cone, varying in the interval $[0,\frac{\pi}{2}]$ and approaching $\frac{\pi}{2}$ in the limit of smooth surface.\footnote{The angle $\O$ actually varies over the full range $[0,\pi]$, but, since the R\'enyi entropy evaluated for a pure state is equal for the region $A$ or for its complement $\bar A$, the function $a_n^{(d)}$ is symmetric for reflections with respect to $\O=\frac{\pi}{2}$, i.e. $a_n^{(d)}(\Omega)=a_n^{(d)}(\pi-\Omega)$ and we can consistently focus on the interval $[0,\frac{\pi}{2}]$}
The function $a_n^{(d)}$ is the universal contribution to the R\'enyi entropy and depends on the angle $\O$ only. Further $\ell$ and $\d$ are the IR and UV regulators, respectively. The former can be thought of as a (macroscopic) length scale characterizing the geometry of the entangling region $A$ (\ie the region enclosed by the twist operator), whereas the latter can be taken to be a short-distance cut-off originating from the infinite number of short-distance correlations in proximity of the twist-operator. The cusp conjecture, in the most general formulation of~\cite{Bueno:2015lza}, states that, for an arbitrary conformal field theory, the leading contribution to $a_n^{(d)}$ for $\O\to \frac{\pi}{2}$ is controlled by the constant $h_n$ introduced in \eqref{eq:oneptT}. Explicitly,
\begin{equation}\label{sigmah}
 a_n^{(d)}(\O) \overset{\scriptscriptstyle \O\to\pi/2}{\sim} 4\,  \s_n^{(d)} (\O-\tfrac{\pi}{2})^2\, ,~~ \sigma_n^{(d)}=\frac{h_n}{n(n-1)}\ \frac{(d-1)(d-2)\,\pi^{\frac{d-4}{2}}\,\Gamma\left[\frac{d-1}2\right]^2}{16\ \Gamma[{d}/{2}]^3}\ \times\,\left\lbrace 
\begin{array}{cll}
\pi && d\,\, \text{odd}\,,\\
1&& d\text{ even}\,.\end{array}\right.
\end{equation}
Restricting to the case $n=1$ and using \eqref{hCTrel}, one finds the following relation between the small angle contribution to the entanglement entropy and the central charge $C_T$ of a CFT
\begin{equation}
 \sigma_1^{ (d)}\equiv\sigma^{ (d)}=C_T\, \frac{\pi^{d-1}(d-1)(d-2)\Gamma[\frac{d-1}{2}]^2}{8\,\Gamma[{d}/{2}]^2\,\Gamma[d+2]}\times\left\lbrace \begin{array}{cll}
\pi&& d \text{ odd}\, ,\\ 
1 && d\, \,\text{even} \,.\end{array}\, \right. \label{ugly}
\end{equation}
In the following, we will apply the theoretical framework introduced in section \ref{sec:twistdef} to this particular deformation and find a connection between $\s_n$ and $C_D$. This allows us to prove the equivalence of the cusp conjecture and eq.~\eqref{eq:CDhrel}.

\subsection{Conical deformation from the displacement operator}

One of the appealing features of the displacement operator is that equation \eqref{deformation} is valid for any kind of deformation of the defect, regardless of whether or not it is smooth. It is then clear that the response \reef{wawa} of the \ren\ entropy to a conical singularity in the limit $\O\to \frac{\pi}{2}$ can be related to the two-point function of the displacement operator \eqref{deformation} integrated over a planar defect with the appropriate profile. In particular combining \eqref{sigman} and \eqref{sigmah}, we obtain
\begin{equation}\label{dispcone}
\frac12 \Sigma^{(d)}\equiv\frac12 \left. \frac{1}{\braket{\tau_n}}\frac{d^2}{d \ep^2} \braket{\tau_n} \right|_{\ep=0}=  4(n-1)\,\s^{(d)}_n \times \left\{\begin{array}{ll}
                          (-1)^{\frac{d+1}{2}}\  \log (\ell/\d) &\qquad d \text{  odd}\\
                          \ (-1)^{\frac{d}{2}}\  \, \ \log^{2\,} (\ell/\d) &\qquad  d \text{  even}
                          \end{array}\right.
\end{equation}
where the first equality is just the definition of $\S^{(d)}$.
In the following, we will compute $\Sigma^{(d)}$ in terms of $C_D$ using \eqref{deformation}. Then, exploiting the conjectured relation \eqref{eq:CDhrel}, we will reproduce the cusp conjecture \eqref{sigmah}.

Consider a planar defect, parametrized by parallel coordinates $x^i$ with $i=3,\ldots,d$, and its deformation into a configuration with a conical singularity at the origin. The two coordinates for the orthogonal directions are $x^a$ with $a=1,2$. To deform the plane into a cone, we introduce spherical coordinates $\{r, \theta^1,\ldots,\theta^{d-3}\}$ in the directions parallel to the entangling surface  and we consider a variation $\ep\,\d x^a$ in the direction $2$ proportional to the radius $r$, \ie
 \begin{equation}
\d x^a=\d^a_2\, r\, .  
 \end{equation}
 Plugging this expression into \eqref{deformation} combined with \eqref{2ptdisp} and using the symmetries of the problem to perform the angular integrations, we are left with
 \begin{equation}
\S^{(d)}
=C_D\,  \O_{d-3}\O_{d-4} \int dr_1\,  dr_2\, \int_0^{2\pi} d \theta_{12} \, \frac{r_1^{d-2}r_2^{d-2} \sin^{d-4} \theta_{12}}{(r_1^2+r_2^2-2r_1r_2 \cos \theta_{12})^{d-1}}.
\end{equation}
 where $\theta_{12}$ is the angle described by the position of the two displacement operators in the plane defined by them and the origin. Further $\O_m = 2\pi^{\frac{m+1}2}/\Gamma(\frac{m+1}2)$ is the volume of a unit $m$-sphere. The integration over $\theta_{12}$ yields
  \begin{multline}\label{cusp2int}
\S^{(d)}
=C_D \frac{2^{d-3} \Gamma \left(\frac{d-3}{2}\right) \Gamma \left(\frac{d-1}{2}\right)}{\Gamma (d-1)} \\ \times\int dr_1\,  dr_2\,\Bigg[ \left| r_1^2-r_2^2\right| ^{-d-1} r_1^{d-2}r_2^{d-2}
\bigg((d-2) r_1^4+2 d\,  r_1^2 r_2^2+(d-2) r_2^4\bigg) \Bigg]
\end{multline}
One has to be particularly careful in the integration over $r_1$ and $r_2$ since we expect a singularity along the line $r_1=r_2$. Therefore it is useful to note the symmetry of the integral under the exchange $r_1\leftrightarrow r_2$ and restrict the integration contour to the region $r_1>r_2$. We then regulate the divergences for $r_1\to r_2$ and for $r_1,r_2\to 0$ with a UV cut-off $\d$, and the divergence for $r_1,r_2 \to \infty$ with an IR cut-off $\ell$. Introducing the variables $x=r_1+r_2$ and $y=r_1-r_2$, the integral takes the form
 \begin{multline}
  \S^{(d)}
=C_D  \frac{2^{3-2d}\pi^{d-2}}{\Gamma
   \left(\frac{d}{2}-1\right) \Gamma \left(\frac{d}{2}\right)} \int_\d^\ell dx\\
   \times\int_{\d}^{x}dy \Bigg[  \left(x^2-  y^2\right)^{d-2} (xy) ^{-d-1} \left((d-1) x^4+2 (d-3) x^2 y^2+(d-1) y^4\right) \Bigg]
 \end{multline}
An additional change of variables $w=(y/x)^2$ yields
\newpage
\begin{multline}
  \S^{(d)}
=C_D  \frac{2^{2-2d}\pi^{d-2}}{\Gamma
   \left(\frac{d}{2}-1\right) \Gamma \left(\frac{d}{2}\right)}\\ \times \int_\d^\ell \frac{dx}{x}  \int_{\left(\frac{\d}{x}\right)^{2}}^{1}dw \Bigg[  \left( 1-  w\right)^{d-2} w ^{-1-\frac{d}{2}}
   \left((d-1) w^2+2 (d-3) w+d-1\right) \Bigg]
   \label{sigmadgen}
 \end{multline}
 Since the treatment of this integral differs substantially in even and odd dimensions, it is convenient to analyze the two cases separately. 
 
 \subsubsection*{Even dimension}
 
 It is useful to note that, for integer $d$, the binomial $(1-w)^{d-2}$ can be converted in a finite sum over powers of $w$. Furthermore, if $d$ is even also the exponent of $w^{1-d/2}$ is an integer, which implies that the integral over $w$ contains a first logarithmic divergence for small $w$. We focus on that contribution and we perform the first integration, which yields 
 \begin{equation}
  \S^{(d)}_{\text{even}}
=C_D\frac{(-1)^{\frac{d}{2}} 2^{5-2d} \pi ^{d-2}    \Gamma \left(d\right)}{d (d-1)\Gamma \left(\frac{d}{2}-1\right) \Gamma
   \left(\frac{d}{2}\right)^3}\int_\d^\ell \frac{dx}{x} \log \frac{x}{\d}  + \cdots
 \end{equation}
 where the missing terms contain power-law divergences.
The last integration can be trivially carried out and the final result is 
 \begin{equation}
  \S^{(d)}_{\text{even}}
=C_D\frac{(-1)^{\frac{d}{2}} 2^{-d} \pi ^{d-\frac{5}{2}}  d\,   \Gamma \left(\frac{d-1}{2}\right)}{\Gamma \left(\frac{d}{2}-1\right) \Gamma
   \left(\frac{d}{2}+1\right)^2} \log^2 (\ell/\d)  + \cdots\,. \label{boat}
 \end{equation}
Comparing this result with eqs.~\reef{ugly} and \reef{dispcone}, we find perfect agreement when using \reef{eq:CDhrel} for $C_D$.

One aspect of the computation deserves a comment: Each of the integrations in eq.~\reef{sigmadgen} contribute one of the logarithmic factors to the final expression \reef{boat}. We can see then that one of the logarithmic singularities arises from $x\sim0$, which corresponds to the region near the tip of the cone (since $x=r_1+r_2$). Further, the second comes from $w\sim0$, which corresponds to the collision of the two displacement operators (since $w\sim r_1-r_2$). Implicitly then, the latter appears everywhere along the entangling surface and is sensitive to the geometry far from the tip of the cone. Of course, this fits in nicely with the lore that in even dimensions, the \ren\ entropy contains a (universal) logarithmic factor that is geometric in nature, \eg see eq.~\reef{4drenyi} below. In a certain sense then, the presence of the cone is completely encoded in the logarithm coming from the integration over $x$ in eq.~\eqref{sigmadgen}, while the second logarithm is  sensitive to the smooth geometry 
far from the tip of the cone and is largely unaware of this singular feature. We also note that $S_n$ may also contain contributions with a single logarithmic factor but these are no longer universal in the presence of the conical singularity \cite{sing}, \eg they will be modified when the cut-off changed because of the logarithm-squared term. 
As we shall see below, similar comments apply for odd dimensions as well. However, the `universal' factor coming from the $w$ integration is simply a constant (independent of $\delta$) and also receives contributions from configurations in which the two displacement operators are separated by a finite distance.

\subsubsection*{Odd dimension}
 
In odd dimensions, it is still true that the binomial $(1-w)^{d-2}$ can be expanded as a finite sum but $1-\frac{d}2$ is not an integer anymore. Hence \reef{sigmadgen} becomes an integral of the form
 \begin{align}
  \S^{(d)}_{\text{odd}}
&=C_D  \frac{2^{2-2d}\pi^{d-2}}{\Gamma
   \left(\frac{d}{2}-1\right) \Gamma \left(\frac{d}{2}\right)} \int_\d^\ell \frac{dx}{x} \sum_{k=0}^{d-2} \binom {d-2}{k} (-1)^k \nonumber\\
   &\times\int_{\left(\frac{\d}{x}\right)^{\frac12}}^{1}dw \left((d-1) w^{k-1-\frac{d}{2}}+2(d-3)w^{k-\frac{d}{2}}+(d-1) w^{k+1-\frac{d}{2}}\right)
 \end{align}
For odd $d$, all the exponents in the last bracket are half-integers, and the integration over $w$ only leads to power-like divergences. The only logarithmic term comes from the integration over $x$, combined with the finite part of the integration over $w$, \ie
\begin{align}
  \S^{(d)}_{\text{odd}}
&=C_D  \frac{2^{2-2d}\pi^{d-2}}{\Gamma
   \left(\frac{d}{2}-1\right) \Gamma \left(\frac{d}{2}\right)} \log(\ell/\d)\\
  &\times\sum_{k=0}^{d-2} \binom {d-2}{k} (-1)^k
  \left(\frac{d-1}{k-\frac{d}{2}} +2\frac{d-3}{k+1-\frac{d}{2}}+\frac{d-1}{k+2-\frac{d}{2}}\right)+\cdots\,.
 \end{align}
 Performing the finite sums, we find
 \begin{equation}
  \S^{(d)}_{\text{odd}}
=C_D\frac{(-1)^{\frac{d+1}{2}} 2^{-d} \pi ^{d-\frac{3}{2}}  d\,   \Gamma \left(\frac{d-1}{2}\right)}{\Gamma \left(\frac{d}{2}-1\right) \Gamma
   \left(\frac{d}{2}+1\right)^2} \log (\ell/\d)  + \cdots\,.
 \end{equation}
Again, substituting for $C_D$ using \reef{eq:CDhrel} produces precise agreement with eqs.~\reef{ugly} and \reef{dispcone}.

\subsection{Wilson lines in supersymmetric theories and entanglement in $d=3$} \label{wilson}

The relation between the expectation value of the stress tensor and the two-point function of the displacement operator has been explored, in fact, at least in one other example of a defect CFT, \ie for Wilson lines \cite{Lewkowycz:2013laa}. In that context, $C_D$ is better known as the Bremsstrahlung function. Indeed, a sudden acceleration of a charged source creates a cusp in the Wilson line that describes its trajectory, and it can be shown that the coefficient of the two-point function of the displacement operator measures the energy emitted in the process \cite{Correa:2012at}. The precise relation between the two quantities is 
\begin{equation}
 C^{WL}_D= 12\, B\,,
\end{equation}
where $B$ is the Bremsstrahlung function. The authors of \cite{Lewkowycz:2013laa} observed that the ratio between $B$ and $h$ (the conformal dimension of the Wilson line) is theory dependent. However, a restricted form of universality is valid within a certain class of conformal gauge theories, whose Bremsstrahlung function is related to the one-point function of the stress tensor through a coefficient that only depends on the dimension of spacetime. This class includes theories with $\mathcal{N}=4$ \cite{Lewkowycz:2013laa} and four-dimensional $\mathcal{N}=2$ \cite{Fiol:2015spa,Mitev:2015oty} supersymmetry. In particular, in three dimensions, the general formula conjectured in \cite{Lewkowycz:2013laa} yields
\begin{equation}
C^{WL}_D = 24\, h^{WL}, \label{3drel}
\end{equation}
where $h^{WL}$ is the constant entering the one-point function of the stress-tensor in the presence of a Wilson line. 

Now the three-dimensional case is especially interesting for us, because twist operators become one-dimensional line operators as well. Furthermore, if we consider holographic CFTs, the calculation of the Wilson line \cite{JJ1,sjrey} and the Ryu-Takayanagi prescription \cite{Ryu:2006bv,Ryu:2006ef} for holographic entanglement entropy both reduce to evaluating the area of extremal surfaces anchored on the AdS boundary. The only difference in the two calculations is the overall factor multiplying the extremal area in evaluating the final physical quantity, but this constant factor will cancel out in the ratio between $C_D$ and $h$. 
Hence for theories which possess a holographic dual and belong to the class for which \eqref{3drel} is valid, \eg ABJM  theory \cite{abjm}, the relation between $\partial_n C_D|_{n=1}$ and $\partial_n h_n|_{n=1}$ has to coincide with \eqref{3drel} --- at strong coupling. Hence it is a nontrivial check that, indeed, formula \eqref{eq:CDhrel} reduces to \eqref{3drel} for $d=3$. Let us make two additional remarks: This agreement is better than required in that $C_D(n) = 24\, h_n$ for all $n$ whereas our argument only indicated a match in the $n\to1$ limit.  Notice, furthermore, that both eqs.~\eqref{eq:CDhrel} and \eqref{3drel} are independent of the coupling. Hence this special relation between the CFT data for the two separate physical observables, \ie Wilson lines and \ren\ entropies, which are apparently unrelated, not only agree at strong coupling but also 
at any coupling.

\section{Entanglement entropy and anomalies in 4d Defect CFTs}
\label{sec:fourd}

In any even number of dimensions, the universal contribution to the R\'enyi entropy \eqref{renyi} depends only on the shape of the spatial region $A$ through local geometric quantities. In four dimensions, in particular, when the theory is conformal, Weyl invariance fixes the universal contribution up to three functions of $n$. If we denote by $\ell$ a characteristic length scale of the entangling surface $\Sigma$, then the Renyi entropy takes the form\footnote{In what follows we suppress the well-known `area law' $\sim (\mu\ell)^2$. Its coefficient is scheme dependent and thus non-universal. In particular, it vanishes within dimensional regularization scheme which we employ throughout this paper. }
\beq
 S_n=\(-{f_a(n)\over 2\pi}\int_{\Sigma} R_\Sigma-{f_b(n)\over 2\pi}\int_{\Sigma} \tilde K_{ij}^a \tilde K_{ij}^a
 +{f_c(n)\over 2\pi}\int_{\Sigma} \gamma^{ij}\gamma^{kl}C_{ikjl}\) \log(\mu \ell) + \lambda_n~,
 \label{4drenyi}
\eeq
where $\gamma^{ij}$ is the inverse of the induced metric on the entangling surface, $\mu$ is an arbitrary mass scale typically chosen to be of order of the inverse cut-off, and $\lambda_n$ is a non-universal constant. Further, $\tilde K_{ij}^a$ is a traceless part of the extrinsic curvature of $\Sigma$
\beq
 \tilde K_{ij}^a= K_{ij}^a - {K^a\over 2} \gamma_{ij}   ~,
\eeq
with $K^a=\gamma^{kl}K_{kl}^a$. Now, two of the coefficients appearing in \eqref{4drenyi} are conjectured to be equal to each other \cite{Lee:2014xwa}:
\beq
f_b(n)=f_c(n)\,.
\label{fconj}
\eeq
This relation has been proven for $n=1$, but remains an open question in general. On the other hand, from our defect CFT point of view, the expression \eqref{4drenyi} has the form of a conformal anomaly, which simply arises because the presence of a defect in the vacuum provides additional ways to violate Weyl invariance. Since the $a$ and $c$ coefficients of the trace anomalies in a generic even dimensional CFT appear in correlation functions of the stress tensor, one might wonder if the same happens in a defect CFT. In this section we show that this is indeed the case, in the sense that $f_b$ and $f_c$ are directly related to $C_D$ and $h$, respectively, \ie
\beq
f_c(n) = {3\pi\over 2} {h_n\over n-1}\,  , \qquad f_b(n) = { \pi^2\over16} {C_D(n) \over n-1}~.
\label{hcdmatch}
\eeq
The relation between $f_c$ and $h_n$ was recently found in the context of entanglement entropy \cite{aitor7}, but both equalities turn out to be true in a generic defect CFT.\footnote{See for instance \cite{Drukker:2008wr} for a discussion of anomalies in the context of surface operators in $\mathcal{N}=4$ SYM. The relations reported in eq. \eqref{hcdmatch} clearly apply for those defects as well.} In the case of the replica defect, they also establish the equivalence of the conjecture \eqref{fconj} with the four-dimensional version of eq. \eqref{eq:CDhrel}. As a first step towards eq. \eqref{hcdmatch}, we notice that by dimensional analysis (or direct calculation), we have
\beq
 \mu{\del\over \del\mu} S_n - \ell {\del\over \del \ell} S_n=0 \quad \Leftrightarrow  \quad  \mu{\del\over \del\mu} S_n = S_n^{\text{univ}} ~,
 \label{main}
\eeq
where $S_n^{\text{univ}}$ denotes the universal Renyi entropy
\beq
 S_n^{\text{univ}}= -{f_a(n)\over 2\pi}\int_{\Sigma} R_\Sigma-{f_b(n)\over 2\pi}\int_{ \Sigma} \tilde K_{ij}^a \tilde K_{ij}^a
 +{f_c(n)\over 2\pi}\int_{\Sigma} \gamma^{ij}\gamma^{kl}C_{ikjl} ~.
 \label{univren}
\eeq

Varying both sides of \reef{main} with respect to the metric and using \reef{sn}, yields
\bea
   {1 \over 1-n} \, \mu{\del\over \del\mu} 
    \sum_\mt{m} \Big( \langle T^{\mu\nu}_\mt{(m)} (x) \rangle_n -  \langle T^{\mu\nu} (x) \rangle_1 \Big) &=& {- 2\over \sqrt{g(x)}} {\delta S_n^{\text{univ}} \over \delta g_{\mu\nu}(x)}
   ~,
 \label{vevT}
   \\
   {1 \over 1-n} \, \mu{\del\over \del\mu} 
   \Big( \sum_\mt{l,m}\langle T^{\mu\nu}_\mt{(l)} (x) T^{\al\bt}_\mt{(m)}(y) \rangle_n-n\langle T^{\mu\nu} (x) T^{\al\bt}(y)\rangle_1 \Big) 
   &=&  {4\over \sqrt{g(y)} } {\delta\over \delta g_{\al\bt}(y)} { 1 \over \sqrt{g(x)} } {\delta S_n^{\text{univ}} \over \delta g_{\mu\nu}(x)} ,
   \non
   \label{vevTT}
\eea
where indices $m$ and $n$ run over the replicas. 
In the next subsection, we build on eq. \eqref{vevT} to prove that $f_c$ appears in the one-point function of the stress-tensor, while eq. \eqref{vevTT} will be needed in subsection \ref{subsec:fbcd} to match $f_b$ with the two-point function of the displacement operator. 

\subsection{$f_c$ and the expectation value of the stress tensor}

Substituting $d=4$ into eq.~\reef{eq:oneptT}, the nontrivial terms in the one-point function of the stress tensor become\footnote{For convenience in this section, we work with the total energy-momentum tensor of the replicated CFT: $T_\mt{tot}^{\m \nu }=\sum_{{\rm m}=1}^nT_\mt{(m)}^{\m \nu }$.}
\bea
 \langle T^{ij}_\mt{tot} \rangle_n&=& - {h_n\over 2\pi } \, {\delta^{ij} \over r^4} + \ldots ~,
 \non
 \langle T^{ab}_\mt{tot} \rangle_n&=&  {h_n\over 2\pi } \, {3 \, \delta^{ac} \, r^2 - 4 \, x^a x^c \over r^{6}} + \ldots ~,
 \label{twist}
\eea
where as usual the indices $a,c$ and $i,j$ denote the two transverse directions and two parallel directions to the entangling surface, respectively. Further, $r$ denotes the transverse distance from the defect with $r^2={\delta_{ac}\, x^a x^c}$.
Note that $h_n$ in the above expression is a constant, \ie we are in the regime when the surface and the background are flat and thus all curvatures can be ignored. While eq.~\reef{eq:oneptT} was written for a planar twist operator, this expression also coincides with the leading singularity for general entangling surfaces if $x$ is sufficiently close to $\Sigma$. In particular, the same constant appears for the conformal weight $h_n$ independently of the geometry of the entangling surface.

Of course, \reef{twist} is independent of $\mu$, and thus one might think that we reached a contradiction with \reef{vevT}. However, this conclusion is too fast. The right hand side of \reef{vevT} vanishes unless $r=0$, but $r=0$ corresponds to a singular point of \reef{twist}. This singularity should be carefully defined as distribution. As we will see, this results in a dependence on a mass scale $\mu$.

In what follows we use dimensional regularization and expand all the results around $d=4$. In particular, we start from the analog of \reef{twist} with dimension of the entangling surface being fixed (\ie two in our case), while the transverse space to the entangling surface is assumed to have dimension $d-2$ (rather than two, as in four dimensions). Hence, the analog of \reef{twist} reads
\bea
 \langle T^{ij}_\mt{tot} \rangle_n&=& - {h_n\over 2\pi } \, {\delta^{ij} \over r^d} ~,
  \label{twist2}
\\
 \langle T^{ab}_\mt{tot} \rangle_n&=&  {h_n\over 2\pi } \,{1\over d-3} \, {3 \delta^{ac} \, r^2 - d \, x^a x^c \over r^{d+2}} 
 ={h_n\over 2\pi (d-2)(d-3)} \( \delta^{ac} \del^2_\perp - \del^a\del^c \) {1\over r^{d-2}} ~,
\nonumber
\eea
where $\del^2_\perp=\delta^{ac}\del_a\del_c$ is Laplace operator in the transverse space.

Now using the standard Fourier integral
\beq
 \int {d^{d-2} k \over (2\pi)^{d-2}} {e^{i k\cdot x} \over (k^2)^\al} = {\Gamma(d/2-\al-1)\over (4\pi)^{(d-2)/2} \Gamma(\al)} \({4\over x^2}\)^{d/2-\al-1} ~,
\eeq
we deduce
\beq
 {1\over (r^2)^{d/2-\al-1}} = {4^\al \pi^{(d-2)/2} \Gamma(\al)\over \Gamma(d/2-\al-1)} (-\del^2_\perp)^{-\al} \delta_\Sigma ~,
 \label{distrib}
\eeq
where equality holds between the distributions and we recall that $\delta_\Sigma$ denotes the delta function in the transverse space with support on $\Sigma$. Now examining the cases $\al=-1+\epsilon$ and $\al=0+\epsilon$ with $\epsilon\ll 1$, and replacing $(-\del^2_\perp)^\epsilon \to \mu^{2\epsilon}$ yields
\bea
 {1\over r^{d-2}}&=&  - \Omega_{d-3} \( { 1 \over 2 \epsilon} + \log(\mu\, r) + \ldots \) \delta_{\Sigma}  ~,
 \non
 {1\over r^d}&=&  - {\Omega_{d-1}\over 4\pi}  \( { 1 \over 2 \epsilon} + \log(\mu\, r) + \ldots \) \del^2_\perp \delta_{\Sigma} ~.
\eea
where $\Omega_{d-1}=2\pi^{d/2}/\Gamma(d/2)$ and ellipses correspond to a finite $\mu$-independent constant as $\epsilon\to 0$.
Consequently, $r^{-d}$ and $r^{-(d-2)}$ although defined by analytic continuation in $d$ are singular when $d = 4$. Hence, to define \reef{twist2} as a sensible distribution, one has to subtract the singular part,
\bea
 \mathcal{R}{1\over r^{d-2}}&=&  - \Omega_{d-3} \(  \log(\mu\, r) + a \) \delta_{\Sigma}  ~,
 \non
 \mathcal{R} {1\over r^d}&=&  - {\Omega_{d-1}\over 4\pi}  \( \log(\mu\, r) + a \) \del^2_\perp \delta_{\Sigma} ~,
\eea
with $a$ an arbitrary constant (which may be absorbed into $\mu$). Note that such a subtraction modifies \reef{twist2} in the limit of coincident points only. Furthermore, the details of this subtraction are not important as long as the result is used in \reef{vevT}
\bea
 -{h_n  \over 4(n-1)} \delta^{ij} \del^2_\perp \delta_{\Sigma}  &=& - 2 {\delta S_n^{\text{univ}} \over \delta g_{ij}(x)}\Big|_{g_{\mu\nu}=\delta_{\mu\nu}} ~,
 \non
 {h_n\over 2(n-1)} \( \delta^{ac} \del^2_\perp - \del^a\del^c \) \delta_{\Sigma}  &=& - 2 {\delta S_n^{\text{univ}} \over \delta g_{ab}(x)}\Big|_{g_{\mu\nu}=\delta_{\mu\nu}}~.
 \label{h_n}
\eea

Next we use \reef{univren} to evaluate the variation on the right hand side. We start from noting that the term proportional to $f_a(n)$ is topological, and therefore its variation vanishes. Hence, in general we need only vary $f_b(n)$ and $f_c(n)$ terms. Now in four dimensions, the following relations hold
\bea
 C^\lambda_{~\,\mu\sigma\nu} &=& R^\lambda_{~\,\mu\sigma\nu}
 - \( g^\lambda_{[\sigma} R_{\nu]\mu}  - g_{\mu [\sigma} R^\lambda_{\nu]} \)
 +{1\over 3} R\, g^\lambda_{[\sigma} g_{\nu]\mu}  ~,
 \non
 \gamma^{ij} \gamma^{kl} C_{ikjl}&=& \gamma^{ij} \gamma^{kl} R_{ikjl} - \gamma^{ij} R_{ij} + {1\over 3} R 
 \non
 &=&{1\over 3} \( \gamma^{ij} \gamma^{kl} R_{ikjl} - \gamma_{ij} g^{\perp}_{\mu\nu} R^{i\mu j \nu} +  g^{\perp}_{\mu\nu} g^{\perp}_{\al\bt}  R^{\mu\al\nu\bt}  \)~,
\eea
where $g^{\perp}_{\mu\nu} = n^a_\mu n^c_\nu\delta_{ac}$ is the metric in the transverse space to $\Sigma$, \ie $g_{\mu\nu}=\gamma_{\mu\nu}+g^{\perp}_{\mu\nu}$. One can use the Gauss-Codazzi  relation
\beq
  \gamma^{ij} \gamma^{kl} R_{ikjl} = R_\Sigma + K_{ij}^a  K^{ij}_a - K^a K_a ~,
\eeq
where $R_{\Sigma}$ is the intrinsic curvature of the entangling surface, to write
\beq
 \gamma^{ij} \gamma^{kl} C_{ikjl}
 ={1\over 3} \( R_\Sigma + K_{ij}^a  K^{ij}_a - K^a K_a - \gamma_{ij} g^{\perp}_{\mu\nu} R^{i\mu j \nu} +  g^{\perp}_{\mu\nu} g^{\perp}_{\al\bt}  R^{\mu\al\nu\bt}  \)
 \label{Weyl}
\eeq

Now recall that \reef{twist2} is valid in the limit when all curvatures (extrinsic, intrinsic and background) are negligibly small. Hence, we expand the relevant curvature components around the flat space, $g_{\mu\nu}=\delta_{\mu\nu}+h_{\mu\nu}$,
\bea
 \delta R_{~~ia}^{ia}&=&{1\over 2} \( 2 \,\del^a\del^i h_{ai} - \del_\perp^2 h_{~i}^i - \del^i\del_i h^a_{~a} \) + \mathcal{O}(h^2) ~,
\non
 \delta R_{~~ab}^{ab}&=&  \del^a\del^b h_{ab} - \del_\perp^2 h^{a}_{~a}  + \mathcal{O}(h^2) ~,
  \label{vars1}
\eea
where we used the results listed in Appendix \ref{varmet} and summation over the repeated indices is assumed.

Next we again use the fact that the integral of the intrinsic curvature over a two-dimensional manifold is a topological invariant, and therefore its variation vanishes. As a result, we obtain\footnote{Note that the third term in \reef{vars1} is a total derivative.}
\bea
 - 2 {\delta S_n^{\text{univ}} \over \delta g_{ij}(x)}\Big|_{g_{\mu\nu}=\delta_{\mu\nu}}&=& - {f_c(n)\over 6\pi} \delta^{ij} \del^2_\perp \delta_\Sigma ~,
 \non
  - 2 {\delta S_n^{\text{univ}} \over \delta g_{ab}(x)}\Big|_{g_{\mu\nu}=\delta_{\mu\nu}}&=&  -{f_c(n)\over 3\pi} \(\del^a\del^b - \delta^{a b} \del^2_\perp \)\delta_\Sigma ~,
  \label{varSn}
\eea
where we have used the following identities
\beq
 {\delta g_{\al\bt}(y) \over \delta g_{\mu\nu}(x)} = \delta^\mu_{(\al}\delta^\nu_{\bt)} \delta (x-y) ~, \quad \int_\Sigma\delta(x-y)=\delta_\Sigma(x_a) ~\text{for}~ y\in\Sigma ~.
\eeq

Comparing \reef{h_n} and \reef{varSn}, yields
\beq
 f_c(n) = {3\pi\over 2} {h_n\over n-1} ~.
\eeq
In full agreement with the existing results for free fields \cite{safdi1}. As we mentioned above, this result was also found with a complementary argument in \cite{aitor7}.

\subsection{$f_b$ and the two-point function of the displacement operator}
\label{subsec:fbcd}

We now turn to the second equality in eq.~\eqref{hcdmatch}. Since we like to find the appearance of $C_D$, we begin by considering the following Ward identity derived in appendix \ref{sec:ward}
\beq
\langle D^a(x) D^c(y) \rangle_n \delta_\Sigma(x) \, \delta_\Sigma(y)  = 
- \langle \nabla_\mu T_{\mt{tot}}^{\mu a} (x) \nabla_\nu T_{\mt{tot}}^{\nu c}(y) \rangle_n \quad \text{for} \quad x\neq y\in\Sigma ~.
\label{defDD}
\eeq
Of course, when either $x$ or $y$ are away from $\Sigma$, the correlator on the right hand side vanishes identically. However, as we will see, it does not vanish when $x$ and $y$ hit the entangling surface $\Sigma$.  This is why the $\delta_\Sigma$'s are explicitly included on the left hand side of the above identity. 
In particular, we are interested in the leading order singularity of $\langle D^a(x) D^c(y) \rangle_n$ when $x$ approaches $y$. In this limit curvature corrections are subleading, \ie both the entangling surface and the background can be regarded as flat. 
From \reef{vevTT}, we have
\bea
{1 \over 1-n} \, \mu{\del\over \del\mu} 
     \langle \del_\mu T^{\mu a}_\mt{tot} (x) ~ \del_\nu T^{\nu c}_\mt{tot}(y) \rangle_n \Big|_{g_{\mu\nu}=\delta_{\mu\nu}}
   &=&   \,  {\del^2 \over \del y^\nu \,\del x^\mu}\left( \, 4{\delta^2 S_n^{\text{univ}} \over \delta g_{\nu c}(y) \delta g_{\mu a}(x)} \Big|_{g_{\mu\nu}=\delta_{\mu\nu}} \right. 
   \non
   &-&\left.2 \, \delta^{\nu c} \, \delta(x-y)\, {\delta S_n^{\text{univ}} \over \delta g_{\mu a}(x)}\Big|_{g_{\mu\nu}=\delta_{\mu\nu}}\right)~.
   \label{DD}
\eea
The results of appendix \ref{varmet} yield\footnote{As before summation over the repeated indices is assumed.}
\begin{equation}
\begin{split}
\delta^2 R^{ia}{}_{ia} =&
 \frac{1}{4} \left( 2\partial_a h_{i\mu} \partial^i h^{a\mu} +\partial_a h_{i\mu} \partial^a h^{i\mu}-2\partial_a h_{i\mu} \partial^\mu h^{ai}+\partial_i h_{a\mu} \partial^i h^{a\mu}\right. \\
 &\left.-2\partial_i h_{a\mu} \partial^\mu h^{ai}+\partial_\mu h_{ai} \partial^\mu h^{ai} \right) \\
&-\frac{1}{4}\left(4\partial^a h_{a\mu} \partial_i h^{i\mu}-2\partial_a h^{a\mu} \partial_\mu h^i_i-2\partial_\mu h_a^a \partial_i h^{i\mu}+\partial_\mu h_a^a \partial^\mu h^i_i\right) \\
&+\frac{1}{2} h^{ab}\left(\partial_a\partial_b h^i_i-2\partial_i\partial_b h^i_a
+\partial_i\partial^i h_{ab}\right) \\
&+\frac{1}{2}  h^{ij}\left(\partial^a\partial_a h_{ij}-2\partial_a\partial_j h^i_a
+\partial_i\partial_j h_a^a\right) \\
&+ h^{ai}\left(\partial_a\partial_i h^j_j-\partial_a\partial_j h_i^j-\partial_i\partial_j h_a^j
+\partial^j\partial_j h_{ai}\right).
\end{split}
\label{riaia}
\end{equation}
Similarly,
\begin{equation}
\begin{split}
\delta^2 R^{ab}{}_{ab} =&
 \frac{1}{2} \left( \partial_a h_{b\mu} \partial^b h^{a\mu} +\partial_a h_{b\mu} \partial^a h^{b\mu}-\partial_a h_{b\mu} \partial^\mu h^{ab}-\partial_b h_{a\mu} \partial^\mu h^{ab}
+\frac{1}{2}\partial_\mu h_{ab} \partial^\mu h^{ab} \right) \\
 &-\partial^a h_{a\mu} \partial_b h^{b\mu}+\partial_a h^{a\mu} \partial_\mu h^b_b
-\frac{1}{4} \partial_\mu h^a_a\partial^\mu h^b_b \\
&+h^{ab} \left(\partial_a\partial_b h^c_c-2 \partial_b\partial_c h^c_a+\partial_c\partial^c h_{ab}
\right) \\
&-2h^{ai}\left(\partial_a\partial_b h^b_i-\partial_b\partial^b h_{ai}-\partial_a\partial_i h^b_b+\partial_i\partial_b h^b_a\right).
\end{split}
\label{rabab}
\end{equation}
and
 \bea
 \delta^2\(K^a_{ij} K_a^{ij}\) \Big|_{g_{\mu\nu}=\delta_{\mu\nu}} &=& 
 {1\over 2} \del_i h_{aj} \del^i h^{aj} + {1\over 2}\del_j h_{ai} \del^i h^{aj} -  \del_a h_{ij}  \del^i h^{aj}
 +{1\over 4} \del_a h_{ij} \del^a h^{ij} ~,
 \non
 \delta^2\(K^a K_a \)\Big|_{g_{\mu\nu}=\delta_{\mu\nu}} &=& \del_i h^{ai} \del_j h^j_a - \del_i h^{ai} \del^a h^j_j + {1\over 4} \del_a h^i_i \, \del^a h^j_j ~.
\eea

These expansions together with \reef{vars1} are sufficient to evaluate the variation on the right hand side of \reef{DD}. There is, however, a significant simplification if we notice that the general term of this variation contains: two delta functions, $\delta_\Sigma$, which restrict the final answer to the entangling surface, one delta function intrinsic to the entangling surface and 4 derivatives, $\del_a$ and $\del_i$, which act on these delta functions. Among all such terms only those with four derivatives parallel to the entangling surface will contribute to the leading singularity of $\langle D^a(x) D^b(y) \rangle_n$ as $x$ approaches $y$. Hence, the relevant part of the variations are
\bea
 \delta^2 R^{ia}{}_{ia} &=& \frac{1}{2}\left( \partial_i h_{aj} \partial^i h^{aj} -\partial_i h_{aj} \partial^j h^{ai} \right)+h^{ai} \left(-\partial_i\partial_j h_a^j
+\partial^j\partial_j h_{ai}\right)+\dots
 \non
  \delta^2\(K^a_{ij} K_a^{ij}\) \Big|_{g_{\mu\nu}=\delta_{\mu\nu}} &=& {1\over 2} \( \del_i h_{aj} \del^i h^{aj} + \del_j h_{ai} \del^i h^{aj}\) + \ldots ~,
  \non
  \delta^2\(K^a K_a \)\Big|_{g_{\mu\nu}=\delta_{\mu\nu}} &=& \del_i h^{ai} \del_j h^j_a  + \ldots
\eea
where the ellipses encode terms which do not contribute to the leading singularity of $\langle D^a(x) D^b(y) \rangle_n$ as $x$ approaches $y$.

Now it follows from \reef{Weyl} that the term proportional to $f_c(n)$ in \reef{univren} does not contribute to the leading singularity of $\langle D^a(x) D^b(y) \rangle_n$ while $f_b(n)$ gives
\beq
 4 \,  {\del^2 \over \del y^\nu \,\del x^\mu} \, {\delta^2 S_n^{\text{univ}} \over \delta g_{\nu c}(y) \delta g_{\mu a}(x)} \Big|_{g_{\mu\nu}=\delta_{\mu\nu}} = 
 - {f_b(n) \over 2\pi} \, \delta^{ac} \, (\del_i\del^i)^2 \delta_\|(x-y) \, \delta_\Sigma(x) \, \delta_\Sigma(y) ~,
\eeq
where $\delta_\|(x-y)$ is the delta function intrinsic to $\Sigma$. Substituting into \reef{DD} and using \reef{defDD}, yields\footnote{It follows from \reef{varSn} that the first variation of $S_n^\mt{univ}$ does not have the same singularity structure as $\langle D^a D^b\rangle$, and therefore it does not contribute.}
\beq
 {1\over 1-n}\mu {\del  \over \del \mu} \langle D^a(x) D^b(y) \rangle_n = {f_b(n) \over 2\pi} \, \delta^{ab} \, (\del_i\del^i)^2 \delta_\|(x-y)  \quad \text{for}
 \quad x,y\in\Sigma~.
\eeq

Now let us recall that the leading singularity of $\langle D^a(x) D^b(y) \rangle_n$ is entirely fixed based on translation invariance along the flat entangling plane and scaling dimension of $\del_\mu T^{\mu\,a}$, \ie up to a constant $C_D$, we have
\beq
 \langle D^a(x) D^b(y) \rangle_n = C_D {\delta^{ab} \over |x-y|^6} = C_D \, { \delta^{ab} \over (x-y)^{2(d-1)} }  \quad \text{for}
 \quad x,y\in\Sigma~.
 \eeq
In particular, we should use the analog of \reef{distrib} to interpret this correlator in the limit $x\to y$.\footnote{The analog is obtained by replacing $\delta_\Sigma$ and $\del_\perp^2$ with $\delta_\|$ and $\del^i\del_i$ respectively.} The final answer takes the form
\beq
 {1\over r^6} =  - {\pi \over 32} \( { 1 \over 2 \epsilon} + \log(\mu\, r) + \ldots \) (\del_i\del^i)^2 \, \delta_\|(r)  ~.
\eeq
Combining altogether, yields
\beq
 {C_D\over n-1} = {16\over \pi^2} \, f_b(n) ~.
\eeq
Further, let us note that this result is in full agreement with \reef{CDhrelder} since $f_b(1)=c= \pi^4\, C_T/ 40$.

\section{Twist operators and the defect CFT data}
\label{sec:singularity}

In the most general sense, a conformal field theory is defined by a set of data, whose knowledge is sufficient to compute all the observables in the theory. A minimal definition of the CFT data includes the spectrum of scaling dimensions of local operators and the OPE coefficients which regulate their fusion. Knowledge of such a set of numbers is sufficient to compute correlation functions with any number of points. However, one might argue that a more complete definition of the CFT data should include those associated to non-local probes, \ie defects: certainly, they are part of the observables of a theory. A possible objection is that the set of defects that can be inserted in a higher dimensional conformal field theory may be very large, even nearly as large as the set of lower dimensional conformal field theories. We may point out that in two dimensions the study of boundaries and interfaces has uncovered a beautiful and simple picture --- see \eg \cite{Cardy:1989ir,Cardy:1991tv,Behrend:1999bn,
Quella:2002ct,Quella:2006de}. However, even in $d=2$, a complete classification of the defect lines which can be placed in a given CFT is a difficult problem, without a solution for the generic case. The situation is better in the special case of topological defects \cite{Frohlich:2006ch}, which have been classified for the Virasoro minimal models \cite{Petkova:2000ip} and for the free boson \cite{Fuchs:2007tx}. In higher dimensions, 
it is perhaps better to think of a theory with a defect as a separate problem, more similar in spirit to the question of which new fixed points can be obtained by coupling two CFTs together. The CFT data that describes a defect CFT are then again associated to correlation functions of local operators in this system, and therefore to the spectrum of primaries and their fusion rules. As we mentioned in section \ref{sec:twistdef}, the main news in the defect CFT setup are given by the spectrum of defect operators, and by the existence of a defect OPE, again regulated by a set of OPE coefficients. 

It is then natural to ask what is the set of CFT data which characterizes the twist operator. This question is not only a simple curiosity. The definition of the replica defect is through a boundary condition in the path-integral. This is often sufficient, but a new definition in terms of CFT data would apply to any conformal field theory, irrespectively of the availability of a path-integral description.\footnote{We thank Davide Gaiotto for a discussion on this point.} Again, some care is needed in setting up this question. The large majority of the OPE coefficients appearing in formulae such as \eqref{defopespin} will depend on the theory in which the twist operator is inserted. However, if an unambiguous characterization exists, it should be possible to single out some universal pattern, unique to this defect and independent of the CFT. In fact, in the present paper, we highlighted the presence of two interesting features. First, the CFT data associated to a flat twist operator always includes a spectrum 
of defect primaries with rational spin under rotations around the defect. Second, we have the suggestion that the coefficient of the two-point function of the displacement and the one of the expectation value of the stress-tensor might be constrained to obey eq. \eqref{eq:CDhrel}. Both these facts are theory independent, but both of them are not unique to the twist operator. Codimension-two defects supporting operators with non-integer transverse spin can be easily constructed --- see for instance \cite{Billo:2013jda} --- while the relation~\eqref{eq:CDhrel} is shared by Wilson lines in a class of three-dimensional supersymmetric gauge theories, as we discussed in section \ref{sec:cusp}. 

Of course, we also understand that the latter constraint will only be obeyed within a special class of CFTs.
Eq. \eqref{eq:CDhrel} is nevertheless remarkable, and one might wonder whether it is possible to understand it from the abstract perspective that we are adopting here. In fact, something special does happen in the defect OPE of the stress tensor, when this relation is fulfilled: a certain number of singular contributions to this OPE disappear, as we now show.
The appearance of the displacement operator in the defect OPE of the stress tensor is constrained by Lorentz and scale invariance to take the following form:
\begin{subequations}
\begin{align}
T^{ij}(x) &\sim \dots + \alpha \frac{x_b D^b\,\de^{ij}}{r^2}+
\beta\, x_b\, \pa_i\pa_j D^b+\ga\, x_b\, \de^{ij}\, \pa_k\pa^k D^b+\dots \\
T^{bi}(x) &\sim\dots  + \de\, \frac{x^b x_c\,\pa^i D^c}{r^2} + \ep\, \pa^i D^b +\dots \\
T^{bc}(x)  &\sim \dots+ \zeta\, \frac{x^b x^c x_a D^a}{r^4}
+\eta\,\frac{\de^{bc}x_a D^a}{r^2}+
\la \frac{D^bx^c+D^cx^b}{r^2}+\dots
\end{align}
\label{stressDOPE}%
\end{subequations}
where $r$ denotes the transverse distance from the defect, as usual. The first ellipsis in each line alludes to the identity and to operators which might be lighter than the displacement, and the second ellipsis indicates less-singular contributions, including higher descendants of the displacement itself. Conformal invariance and conservation of the energy-momentum tensor place constraints on the coefficients in eq.~\eqref{stressDOPE}, and only two of them are independent. More interestingly, these two numbers are in fact fixed in terms of the conformal weight $h$ and the coefficient $C_D$. A proof of these statements appears in \cite{billo:confdef}, but it is not difficult to understand how they may come about. The form of the OPE is determined by the two-point function of the displacement operator with the stress-tensor, which is fixed by conformal symmetry up to three coefficients:\footnote{Our conventions differ from the ones in \cite{billo:confdef} in the following way: the displacement has 
opposite sign --- and with it the constants $b_{DT}^i$ --- and the conformal weight of the twist operator is denoted there as $a_T=-d\, h/2\pi$.}
\begin{align}\label{TDcorr}
\braket{D^a(x_1)T^{ij}(x_2)}&= \frac{x_2^a}{(x_{12}^2)^{d-1}r^2}
\frac{1}{d} \left\{b^1_{DT}\left(\frac{4d\, r^2\, x_{12}^i x_{12}^j}{x_{12}^4}-\delta^{ij}\right)+b^2_{DT}\,\delta^{ij}\right\},\notag\\
\braket{D^a(x_1)T^{ib}(x_2)}&=\frac{x_{12}^i}{(x_{12}^2)^d}\left\{ 2b^1_{DT}\, \frac{x^a_2 x_2^b}{r^2} \left(1-\frac{2r^2}{x_{12}^2}\right)+b^3_{DT}\left(\delta^{ab}-\frac{x^a_2 x_2^b}{r^2}\right) \right\},\notag\\
\braket{D^a(x_1)T^{bc}(x_2)}&=\frac{1}{(x_{12}^2)^{d-1}r} \left\{b^1_{DT}\,\frac{x_2^a}{r}\left[ 
\frac{x_2^b x_2^c}{r^2}\frac{\left(x_{12}^2-2 r^2\right)^2}{x_{12}^4}
-\frac{1}{d}\delta^{bc}\right]\right.\notag\\
&+\left.b^2_{DT}\,\frac{x_2^a}{r} \left(\frac{x_2^b x_2^c}{r^2}-\frac{d-1}{d}\delta^{bc}\right) \right.\notag\\
&+\left.b^3_{DT} \left(1-\frac{2r^2}{x_{12}^2}\right)\left(\frac{\delta^{ab}x_2^c+\delta^{ac}x_2^b}{2r}-\frac{x_2^ax_2^b x_2^c}{r^3} \right) \right \}.
\end{align}
Here $x_1$ is, of course, confined to the defect while $r$ denotes the transverse distance of $T^{\m\n}$ from the defect, and $x_{12}^\m=x_1^\m-x_2^\m$. 
Now imagine integrating the displacement along the defect: this is equivalent to an infinitesimal translation in the direction 
labeled by $a$. We see that the integrated two-point function is proportional to the derivative $\pa_a \braket{T^{\m\n}}$, whence two linear relations follow:
\begin{subequations}
\begin{align}
b_{DT}^2 &= \frac{1}{d-1} \left(\frac{d}{2} b_{DT}^3-b_{DT}^1\right), \label{b2}\\
b_{DT}^3 &=  2^{d} d\,\pi^{-\frac{d-1}{2}} \Gamma\! \left(\frac{d+1}{2}\right) \frac{h}{2\pi}.\label{b3}
\end{align}
\label{TDintred}
\end{subequations}
The first of the \eqref{TDintred} reduces to two the independent coefficients and is compatible with conservation. On the other hand, if we contract the same two-point function with a derivative, we obtain the two-point function of the displacement via eq. \eqref{stressward}, and this provides the relation involving $C_D$:
\beq
b_{DT}^2=\frac{1}{d-2}\left(\frac{d}{2}\,b_{DT}^3-2\frac{C_D}{2\pi}\right).
\label{b2b3cd}
\eeq
We see that the two-point function, and so the relative contribution to the defect OPE of the stress-tensor, are fixed in terms of $C_D$ and $h$.

Let us now consider the most singular contributions in every component in \eqref{stressDOPE}. In Euclidean signature, all terms in the the OPE of $T^{bi}$ and $T^{bc}$ have the same degree of singularity. We can still define the most singular terms in Lorentzian signature, by considering a spacelike defect --- this is especially natural when talking about \ren\ entropy. Now as the insertion approaches the null cone, the individual $x^a$ may remain finite while $r$ approaches zero. In this circumstance, the most singular terms are those multiplied by $\alpha,\, \de$ and $\zeta$. Comparing eqs. \eqref{stressDOPE} and \eqref{TDcorr}, we easily find
\beq
\a = \frac{1}{d\,C_D} (b_{DT}^2-b_{DT}^1), \quad
\de = \frac{1}{2(d-1)C_D} (2 b_{DT}^1-b_{DT}^3), \quad
\zeta = \frac{1}{C_D} (b_{DT}^1+b_{DT}^2-b_{DT}^3).
\eeq 
Remarkably, the three constants vanish when eq.~\eqref{eq:CDhrel} holds, \ie
\begin{equation}
 \a=\de=\zeta=0 \iff C_D(n)= d\, \Gamma\left(\tfrac{d+1}{2}\right)\, \left(\tfrac{2}{\sqrt{\pi}}\right)^{d-1}\, h_n
\end{equation}

This observation is appealing, even if its meaning remains somewhat obscure. One may speculate that the twist operator is a ``mild'' defect, in some sense. It is obtained through a modification of the geometry, rather than the addition of local degrees of freedom, and now we see that the OPE of the stress-tensor is less singular than for a generic defect. However, this idea should not be taken too literally. The identity appears in the same defect OPE, with a more severe singularity. Moreover, lighter defect operators with respect to the displacement might exist --- in fact, they do in a free scalar theory, as discussed in Appendix \ref{sec:scalar}. Some of them may also appear in the defect OPE of the stress tensor. Whatever the right interpretation may be, it is worth emphasizing that it would have been probably difficult to recognize the special character of the relation \eqref{eq:CDhrel}, without adopting the defect CFT perspective.

\section{Discussion}
\label{sec:discussion}

Twist operators were originally defined in examining \ren\ entropies
in two-dimensional CFTs \cite{Calabrese:2004eu,Cardy:2007mb} and they are easily understood in this context since they are local primary operators. As discussed in section \ref{sec:twistdef}, twist operators are formally defined for general QFTs through the replica trick, as in eq.~\reef{sigman}. In higher dimensions then, they become nonlocal surface operators and their properties are less well understood.  In the present paper, we have begun to explore twist operators for CFTs in higher dimensions from the perspective of conformal defects. This approach naturally introduces a number of tools that are unfamiliar in typical discussions of \ren\ entropies. In particular, our discussion has focused on the displacement operator $D^a$, which appears with the new contact term in the Ward identity \reef{stressward}. 

A key role of the displacement operator is to implement small local deformations of the entangling surface, as in eq.~\reef{211}. As shown in eq.~\reef{deformation}, the expectation value of the twist operator itself only varies at second order for such deformations of a planar (or spherical) entangling surfaces and is determined by the two-point function \reef{2ptdisp} of the displacement operator. This behaviour was previously seen in holographic studies of the so-called entanglement density \cite{Nozaki:2013vta} and more recently in \cite{new}. These results correspond to the special case of the $n\to1$ limit in eq.~\reef{deformation}. We might also like to note that the connection with Wilson lines in holographic conformal gauge theories discussed in section \ref{wilson} would also relate these entanglement variations to the wavy-line behaviour of Wilson lines \cite{wavy}.

Our main result was to unify a variety of distinct conjectures, summarized at the end of section \ref{sec:twistdef}, about the shape dependence of \ren\ entropy to a constraint \reef{eq:CDhrel} relating the coefficient defining the two-point function of the displacement operator and the conformal weight of the twist operator. While the connections between these conjectures, were already considered in \cite{Bueno:2015lza} --- see also discussion in \cite{new} --- eq.~\reef{eq:CDhrel} appears to provide the root source with a relation between two pieces of CFT data characterizing the twist operators. 

One of these conjectures was the equivalence of the coefficients $f_b(n)$ and $f_c(n)$ appearing in the universal part of the four-dimensional R\'enyi entropy for general $n$ \cite{Lewkowycz:2014jia,safdi1}. However, it was very recently shown that this equivalence does not hold for four-dimensional holographic CFTs dual to Einstein gravity \cite{dong34}. As a consequence, it follows that eq.~\eqref{eq:CDhrel} does not hold for general $n$ in these holographic CFTs either. On the other hand, this relation does hold in the vicinity of $n=1$ for general CFTs. That is, the recent results of \cite{new} demonstrate that the first order expansion of eq.~\reef{eq:CDhrel} about $n=1$ is a constraint which holds for generic CFTs. Despite the fact that this relation does {\it not} hold for all values of $n$ for all CFTs, it is still interesting to ask for precisely which CFTs does this constraint hold. It seems that free field theories are a good candidate for such a theory. The results of \cite{Bueno:2015qya,Dowker1,Dowker2} for the universal corner contribution to the \ren\ entropy in three dimensions imply that eq.~\eqref{eq:CDhrel} holds for free scalars and fermions in this dimension. Further, our calculations in Appendix \ref{sec:scalar}  confirm that it holds for free massles scalars in four dimensions. We hope to return to this question in future work \cite{future}. 

While eq.~\eqref{eq:CDhrel}, and hence the related conjectures, are not completely universal, it is nevertheless a remarkable relation.
It may still be interesting to explore other implications which this relation has for \ren\ entropies in other geometries and other dimensions. 
For example, it could provide a relation (for arbitrary $n$) between different coefficients appearing in the universal contribution to the \ren\ entropy in $d=6$ or higher even dimensions, along the lines of our four-dimensional discussion in section \ref{sec:fourd}. 

Recalling that the twist operator is a local primary in two-dimensional CFTs, we might ask how the displacement operator appears in this context. Here, the natural object is the first descendant, \ie derivative, of the twist operator which would be analogous to the combination of the displacement and twist operators together. This matches the appropriate contact term in the two-dimensional version of the Ward identity \reef{stressward}. Here we refer to an analogy (rather than a precise match) keeping in mind that as a local operator, the two-dimensional twist operator can be moved but not deformed. Still one might make sense of the two-point correlator \reef{2ptdisp} by considering a ``spherical" entangling surface. In two dimensions, the (zero-dimensional) sphere would correspond to two points whose separation defines the diameter of the sphere. Hence eq.~\reef{2ptdisp} would be given by taking derivatives of the correlator of two twist operators and hence one finds that the corresponding $C_D$ is indeed 
proportional to the conformal weight $h_n$.

Our discussion has highlighted $h_n$ and $C_D$ as two pieces of CFT data which characterize twist operators. With this perspective of regarding the twist operator as a conformal defect, we began in section \ref{sec:singularity} to consider the question of what are the defining characteristics of the twist operator? Certainly the relation \reef{eq:CDhrel} would be an important feature since, as we noted there, it has an interesting impact on the defect OPE with the stress tensor. However, this relation is not completely universal and, as described in section \ref{sec:twistdef}, this property is also shared by Wilson line operators in certain superconformal gauge theories. Another important property discussed in section \ref{sec:twistdef} is that the spectrum of defect operators can contain operators with fractional spins $k/n$. Certainly, our analysis of the free scalar theory in appendix \ref{sec:scalar} explicitly reveals the presence of such operators. But again twist operators are not unique in this regard. Another interesting point that arises in our discussion is that the twist operators are naturally defined for integer $n$ but in discussing $h_n$ and $C_D$, as well as the \ren\ entropy, one continues the results to real $n$ almost immediately. Here derivatives of correlators with respect to the \ren\ entropy index are naturally defined in terms of the modular Hamiltonian \cite{Hung:2014npa,solo}. This seems to point to a unique characteristic of twist operators in higher dimensions. In any event, better understanding the definition of the twist operator as a conformal defect remains an open question. Undoubtedly it is a question whose answer will produce a better understanding of the entanglement properties of CFTs, and perhaps QFTs more generally.

\acknowledgments
We would like to thank Marco Bill\`o, J\"urgen Fuchs, Edoardo Lauria, Aitor Lewkowycz, Jonathan Toledo and especially Davide Gaiotto and Vasco Goncalves for valuable comments and correspondence. Research at Perimeter Institute is supported by the Government of Canada through
Industry Canada and by the Province of Ontario through the Ministry of
Research \& Innovation. RCM acknowledges support from an NSERC Discovery grants and funding from the Canadian Institute for Advanced Research. The work of LB is supported by Deutsche Forschungsgemeinschaft in Sonderforschungsbereich 676 ``Particles, Strings, and the Early Universe''. The work of MS is supported in part by the Berkeley Center for Theoretical Physics and by the National Science Foundation (award numbers 1214644, 1316783, and 1521446). 

\appendix

\section{Ward identities in the presence of a twist operator}
\label{sec:ward}

This appendix is devoted to the Ward identities obeyed by the stress tensor in the presence of a twist operator. We shall focus on the displacement operator and opt for a streamlined derivation. We refer to \cite{billo:confdef} for a more detailed account.
Let us consider a $q$-point correlator of the scalar fields on an arbitrary replicated manifold $\M_n$
\beq
 \Gamma(x_1, x_2,\ldots x_q, \Sigma, g_{\mu\nu}) \equiv 
 \langle \phi(x_1) \phi(x_2) \cdots \phi(x_q) \rangle_n \equiv \langle \phi(x_1) \phi(x_2) \cdots \phi(x_q) \, \tau_n(\Sigma) \rangle~,
\eeq 
where $n$ is the replica parameter and $\tau_n(\Sigma)$ is the twist operator associated with the entangling surface $\Sigma$. By definition $\Gamma(x_1, x_2,\ldots x_q, \Sigma, g_{\mu\nu})$ transforms as a scalar under diffeomorphisms of the manifold. This means it will be unchanged if we simultaneously make the following infinitesimal replacements
\bea
 \delta x^\mu_i&=& \xi^\mu |_{x_i} \quad \text{for}\quad i=1,\ldots q ~,
 \non
 \delta x^\mu &=&  (\xi_\al \, n^\al_{\hat c})\,n^\mu_{\hat c} \quad \text{for}\quad x^\mu\in\Sigma~,
 \non
 \delta g_{\mu\nu}&=&-\nabla_\mu \xi_\nu - \nabla_\nu \xi_\mu~,
\eea
where $n_{\hat a}^\mu$ with $a=1,2$ denotes an orthonormal basis of vectors in the transverse space to $\Sigma$. 
Thus to leading order in $\xi^\mu$ we have
\bea
 0&=&-\sum_{i=1}^q \xi^\mu |_{x_i} \langle \phi(x_1)\cdots\del_\mu\phi(x_i)\cdots\phi(x_q)\rangle_n 
 + \langle \phi(x_1) \phi(x_2) \cdots \phi(x_q) \int_\Sigma   \xi^\al D_\al \rangle_n
 \non
 &+& \langle \phi(x_1) \phi(x_2) \cdots \phi(x_q) \int_{\M_n} T^{\mu\nu} \nabla_\mu\xi_\nu \rangle_n ~,
\eea
where $D^\al(y)=n^\al_{\hat c} D_{\hat c}(y)$ is a local operator which implements displacement of the surface operator, $\Phi_\Sigma$, at $y^\mu\in\Sigma$ (analog of $\del_\mu$ for a scalar operator $\phi(x)$). Now recall that $\xi^\mu$ is arbitrary, but must be the same vector on all the sheets in the replicated geometry. With this in mind, we arrive at the following Ward identity
\bea
 0&=&-\sum_{i=1}^q \delta(x-x_i) \langle \phi(x_1)\cdots\del_\nu\phi(x_i)\cdots\phi(x_q)\rangle_n
 + \delta_\Sigma(x) \, \langle \phi(x_1) \phi(x_2) \cdots \phi(x_q)  D_\nu(x) \rangle_n
 \non
 &-&\sum_{m=1}^n \langle \phi(x_1) \phi(x_2) \cdots \phi(x_q) \nabla_\mu T^{\mu}_\nu(x_m) \rangle_n ~,
\eea
where $x_m$ is a point on the $m$-th replica. Of course, this is a more precise expression of the identity `loosely' introduced in eq.~\reef{stressward}.
This Ward identity defines the displacement operator by specifying its matrix elements: the only additional input is the one of locality of the theory and of the defect, which guarantees that the displacement is a local operator.

Next we assume that there are no scalar field insertions and consider a special case when $\xi^\mu$ is peaked around the two given disjoint points $x$ and $y$, but otherwise is arbitrary. Then expanding to linear order in $\xi^\mu$ around $x$ and $y$ and using the above Ward identity, results in (from the cross term $\xi^\mu(x)\xi^\nu(y)$)\footnote{Note that there are two cross terms of the form $\delta_\Sigma(y)\langle \nabla_\mu T^{\mu a}_\mt{tot} (x) D^c(y) \rangle_n $. They vanish identically since only one stress tensor hits the defect, whereas the correlator $\langle T^{\mu a}_\mt{tot} (x) D^c(y) \rangle_n$ for $x\notin\Sigma$ is conserved.}
\beq
\delta_\Sigma(x) \, \delta_\Sigma(y) \, \langle D^a(x) D^c(y) \rangle_n + 
\langle \nabla_\mu T^{\mu a}_\mt{tot} (x) \nabla_\nu T^{\nu c}_\mt{tot}(y) \rangle_n =0  
\quad \text{for} \quad x\neq y ~.
\label{ward2}
\eeq

\section{Small variations of the metric}
\label{varmet}
Consider a small perturbation of the flat metric $g_{\mu\nu} = \delta_{\mu\nu}+h_{\mu\nu}$, then we find the following variations: \\
Christoffel symbols 
\beq
 \delta\Gamma^\lambda_{\mu\nu}={1\over 2} \delta^{\lambda\rho} \( \del_\mu h_{\rho\nu} + \del_\nu h_{\mu\rho} - \del_\rho  h_{\mu\nu}\)
\eeq
Riemann tensor
\bea
 R_{\rho\mu\sigma\nu}&=&{1\over 2} \( \del_{\mu}\del_{\sigma} g_{\rho\nu} + \del_{\rho}\del_{\nu} g_{\mu\sigma} 
 - \del_{\mu}\del_{\nu} g_{\rho\sigma} - \del_{\rho}\del_{\sigma} g_{\mu\nu} \) 
 + g_{\lambda\bt} \( \Gamma^\lambda_{\mu\sigma} \Gamma^\bt_{\rho\nu} - \Gamma^\lambda_{\mu\nu} \Gamma^\bt_{\rho\sigma} \)
 \non
 \delta R_{\rho\mu\sigma\nu}&=&{1\over 2} \( \del_{\mu}\del_{\sigma} h_{\rho\nu} + \del_{\rho}\del_{\nu} h_{\mu\sigma} 
 - \del_{\mu}\del_{\nu} h_{\rho\sigma} - \del_{\rho}\del_{\sigma} h_{\mu\nu} \)  ~,
 \non
 \delta^2 R_{\rho\mu\sigma\nu}&=& \delta_{\lambda\bt} \( \delta\Gamma^\lambda_{\mu\sigma} \delta\Gamma^\bt_{\rho\nu} 
 - \delta\Gamma^\lambda_{\mu\nu} \delta\Gamma^\bt_{\rho\sigma} \)
\eea
Surface forming normal vectors
 \bea
 g^{\mu\nu}  \,n^a_\mu n^c_\nu &=& \delta^{ac} \quad \Rightarrow \quad 
 \delta n^a_\mu n^{c\mu} + n^{a\mu} \delta n^c_\mu = n^{a\mu} n^{c\nu} \delta g_{\mu\nu}  ~,
 \non
 n^a_\mu t^\mu_i &=& 0 \quad~~\, \Rightarrow \quad \delta n^a_\mu t^\mu_i =0 ~,
\eea
where $t^\mu_i=\del x^\mu/\del y^i$ are tangent vectors to the entangling surface. Thus,
\beq
 \delta n^a_\mu =  A^a_{~c} n^c_\mu ~,
\eeq
with 
\beq
 A^1_{~1}={1\over 2}n^{1\mu} n^{1\nu} \delta g_{\mu\nu} ~, \quad A^2_{~2}={1\over 2} n^{2\mu} n^{2\nu} \delta g_{\mu\nu} ~, \quad 
 A^1_{~2}+A^2_{~1} = n^{1\mu} n^{2\nu} \delta g_{\mu\nu}~.
\eeq
Extrinsic curvatures
\beq
 \delta K^a_{ij}=\delta\(\nabla_i n^a_j\)=-\delta\Gamma_{ij}^\mu n^a_\mu + \nabla_i\delta n^a_j=
 -{1\over 2}  n^{a\mu}\(\nabla_i\delta g_{\mu j} + \nabla_j\delta g_{\mu i} - \nabla_\mu \delta g_{ij} \)+A^a_{~c} K^c_{ij} ~.
\eeq
Transverse metric 
\beq
 g^\perp_{\mu\nu}=n^a_\mu n^c_\nu\delta_{ac}  \quad \Rightarrow  \quad 
 \delta g^\perp_{\mu\nu} = A^a_{~b} n^b_\mu n^c_\nu \delta_{ac} +  n^a_\mu A^c_{~b} n^b_\nu  \delta_{ac} ~,
\eeq
or equivalently
\beq
 \delta g^\perp_{\mu\nu} = (n^{1\al} n^{1\bt} \delta g_{\al\bt})n^1_\mu n^1_\nu +(n^{2\al} n^{2\bt} \delta g_{\al\bt})n^2_\mu n^2_\nu
 +(n^{1\al} n^{2\bt} \delta g_{\al\bt})( n^2_\mu n^1_\nu + n^1_\mu n^2_\nu ) ~.
\eeq

\section{Displacement operator for the free scalar}
\label{sec:scalar}
In this appendix, we consider the theory of a free scalar in four dimensions, and we explore the defect OPE of the low lying bulk primaries. In doing so, we give a concrete expression for the displacement operator in terms of Fourier modes of the fundamental field and we verify the conjecture \eqref{eq:CDhrel} for this particular case.
Given the Lagrangian of a four-dimensional free massless boson 
\begin{equation}
 \mathcal{L}=\frac12 (\pa_\mu \phi)^2\,,
\end{equation}
the propagator in presence of a conical singularity with an angular excess $2\pi(n-1)$ placed in $r=0$ can be derived \cite{Guimaraes:1994sw}: 
\begin{equation}\label{eq:prop}
 \braket{\phi(x)\phi(x')}_n=\frac{\sinh(\frac{\eta}{n})}{8\pi^2 n r r' \sinh \eta \left(\cosh(\frac{\eta}{n})-\cos(\frac{\theta}{n})\right)},
\end{equation}
where 
\begin{equation}
 \cosh\eta=\frac{r^2+{r'}^2+y^2}{2r r'}.
\end{equation}
We use alternatively polar coordinates around the defect with $x=(r,\theta,y^1,y^2)$, $x'=(r',0,0,0)$ or complex coordinates $x=(z,\bar z, y^i)$, $x'=(z',\bar z', 0)$ with $z=r e^{i\theta}$ and $z'=r'$. Assuming integer values of $n$ and expanding \eqref{eq:prop} in the defect OPE limit, \ie for $r\to 0$ and $r'\to 0$, one finds
\begin{equation}\label{eq:propexp}
 \braket{\phi(r,\theta,y^i)\phi(r',0,0)}_n=\frac{1}{4 n \pi^2} \left(\frac{1}{y^2} +2 \sum_{k=1}^{n-1} \frac{r^{\frac{k}{n}} {r'}^{\frac{k}{n}}}{(y^2)^{1+\frac{k}{n}}} \cos\frac{k\theta}{n}- \frac{r^2+{r'}^2-2r r' \cos\theta}{y^4} +\cdots\right)
\end{equation}
where $y^2=(y^1)^2 + (y^2)^2$ and the ellipsis indicates terms with higher powers of $r/y$ and $r'/y$.
This result can be precisely reproduced by the following OPE expansion for the field $\phi$ \footnote{The two contributions proportional to $r^2$ and ${r'}^2$ in \eqref{eq:propexp} originate from the descendant $\pa_i\phi$.}
 \begin{equation}\label{phiope}
 \phi(z,\bar z)=\phi(0)+\frac{1}{2\pi \sqrt{n}} \sum_{k\in \mathbb{N}} \left( z^{\frac{k}{n}}  O_{\frac{k}{n}} + \bar{z}^{\frac{k}{n}}  \bar O_{\frac{k}{n}}\right)+\cdots
\end{equation}
where the operators $ O_{\frac{k}{n}}$ are defect primaries with transverse spin $s=\frac{k}{n}$ and scaling dimension $\D=s+1$ and the ellipsis indicates contributions from the descendants. This spectrum of twist-one\footnote{We are calling twist the difference between the scaling dimension and the charge under a transverse rotation. However, let us stress that the latter is a global symmetry from the point of view of the defect theory} defect primaries can be easily understood through the requirement that every conformal family appearing on the r.h.s. of \eqref{phiope} is annihilated by the Laplace operator. Indeed, the latter reduces to the two-dimensional $\partial_z\partial_{\bar{z}}$ differential operator once we disregard descendants, and the holomorphicity property of the contribution of defect primaries to the OPE quickly follows. On the other hand the possible values of the spin are fixed by the symmetry preserved by the defect, \ie a $n$-fold cover of $SO(2)$. The normalization of the operators is 
fixed by 
\begin{equation}
 \braket{O_\frac{k}{n} \bar O_{\frac{k}{n}}}_n=\frac{1}{(y^2)^{1+\frac{k}{n}}}
\end{equation}
Let us make one more comment on the nature of the defect spectrum. The twist operator is responsible for the presence of a tower of primaries with non-integer transverse spin. While these Fourier modes do not possess a local expression in terms of the elementary field, this is not so for the defect operators with integer spin. Their contribution to the defect OPE is modified by the defect, but we can still identify them with derivatives of $\phi$ in directions orthogonal to the defect.\footnote{This is somewhat loose: a defect primary will in general be a combination of derivatives orthogonal and parallel to the defect. The one exception is $\partial_a \phi$, for which no mixing happens.} In particular, it will be important in a moment that a defect operator $O_1=\pa_z \phi$ exists.

We expect to find evidence of the presence of the displacement operator in the defect OPE expansion of the scalar operator $\phi^2$. Therefore we consider the connected correlator 
\begin{equation}\label{eq:corrphi2}
 \braket{\phi(x)^2 \phi(x')^2}_n-\braket{\phi(x)^2}_n\braket{\phi(x')^2}_n=2\, {\braket{\phi(x) \phi(x')}_n}^2
\end{equation}
in the defect OPE limit and we extract the contribution given by operators of dimension $3$ (spin $1$), which reads
 \begin{equation}\label{eq:phi2corr}
 \left.\braket{\phi(x)^2 \phi(x')^2}_n\right|_{\text{spin }1}\sim \frac{r r' \cos \theta}{4 n^2 \pi^4 y^6} (n+1) 
\end{equation}
This formula can be interpreted in terms of the OPE expansion of $\phi^2$, which can be obtained by studying the fusion of two $\phi$ OPEs. In particular at dimension $3$ one has several possible contributions coming from the combination of all the possible spins summing to $1$ and the result is
\begin{equation}\label{eq:OPEphi2}
 \phi^2(z,\bar z)\sim \cdots+\frac{1}{4\pi^2 n}\sum_{k=0}^n \left(z\,  O_{(\frac{k}{n},\frac{n-k}{n})}+\bar z\, \bar O_{(\frac{k}{n},\frac{n-k}{n})}\right) +\cdots
\end{equation}
where $O_{\big(\frac{k}{n},\frac{k'}{n}\big)} = O_\frac{k}{n} O_\frac{k'}{n}$ and the ellipses indicate that we are focusing only on the spin-one contribution. Notice that the sum in \eqref{eq:OPEphi2} is redundant since $O_{\big(\frac{k}{n},\frac{k'}{n}\big)}=O_{\big(\frac{k'}{n},\frac{k}{n}\big)}$, nevertheless we keep this notation so as not to clutter the following expressions. Inserting the OPE in the two-point function and performing the Wick contractions, one obtains
\begin{equation}
 \left.\braket{\phi(x)^2 \phi(x')^2}_n\right|_{\text{spin }1}=\frac{2(z\bar z'+z' \bar z)}{16\pi^4 n^2} \sum_{k=0}^n \braket{O_\frac{k}{n}\bar O_\frac{k}{n}}\braket{O_\frac{n-k}{n}\bar O_\frac{n-k}{n}}=\frac{z\bar z'+z' \bar z}{8 n^2 \pi^4 y^6} (n+1) \,,
\end{equation}
in agreement with \eqref{eq:phi2corr}. The degeneracy we just observed complicates the task of singling out the displacement operator. In the following we will start from a general Ansatz and derive a set of constraints which allows to fix the precise form of the displacement operator for $n\leq 5$ and to extrapolate a general pattern for higher $n$. In the process we will also prove that for this specific theory the relation \eqref{eq:CDhrel} holds for any $n$. 

We start from the general linear combination~\footnote{Here we discuss only the holomorphic part of the displacement operator, but analogous considerations are valid for the anti-holomorphic component $\bar D$.}
\begin{equation}
 D=\frac{1}{2\pi n^2}\sum_{k=0}^n c_{k} O_{(\frac{k}{n},\frac{n-k}{n})}
\end{equation}
where the normalization factor has been introduced for future convenience.
The redundancy of the sum gives the first constraint on the coefficients
\begin{equation}
 c_k=c_{n-k}
\end{equation}

In order to find further constraints we compute the coupling of the displacement with $\phi^2$ and with the stress tensor
\begin{equation}
 T_{\mu\nu}=\pa_\mu \phi \pa_\nu \phi -\frac12 \d_{\mu\nu} \pa \phi\cdot \pa \phi -\frac16 \left(\pa_\mu\pa_\nu -\d_{\mu\nu} \pa^2\right) \phi^2\, .
\end{equation}
The former is fixed by the Ward identity
\begin{equation}\label{eq:Wardphi2}
 \int d^2 y \braket{\phi^2(z,\bar z,0) D(y^i)}_n=\pa_z \braket{\phi^2(z,\bar z,0)}_n
\end{equation}
whereas the latter is determined in terms of $C_D$ and $h_n$ by equations \eqref{TDcorr}, \eqref{b2}, \eqref{b3} and \eqref{b2b3cd}.

We start with the coupling to $\phi^2$. The bulk-defect correlator $\braket{\phi^2 O_{(\frac{k}{n},\frac{n-k}{n})}}_n$ is fixed by symmetry up to a normalization which can be extracted from the OPE \eqref{eq:OPEphi2}. The result is
\begin{equation}
 \braket{\phi^2(z, \bar z,0)\, O_{(\frac{k}{n},\frac{n-k}{n})}(y)}_n=\frac{\bar z}{2\pi^2 n (y^2+z\bar z)^3}
\end{equation}
On the other hand the one-point function $\braket{\phi^2(z,\bar z,0)}_n$ on the r.h.s. of \eqref{eq:Wardphi2} is simply
\begin{equation}
  \braket{\phi^2(z,\bar z,0)}_n=\frac{1-n^2}{48 n^2 \pi^2 z\bar z}
\end{equation}
It is then clear that the Ward identity \eqref{eq:Wardphi2} gives a constraint on the sum of the coefficients $c_k$. Explicitly
\begin{equation}
 \sum_{k=0}^n c_k=\frac{(n-1)n(n+1)}{6}
\end{equation}
Notice that the r.h.s. of this expression is always an integer.

We now move to the computation of the two-point function of the displacement with the stress tensor. By standard Wick contraction one can compute the coupling of $O_{(\frac{k}{n},\frac{n-k}{n})}$ with the parallel components of the stress tensor $T^{ij}$. This gives
\begin{equation}\label{eq:TijO}
 \braket{T^{ij}(z,\bar z,0)\,O_{(\frac{k}{n},\frac{n-k}{n})}(y)}=\frac{2k(n-k)}{n^3 \pi^2} \frac{\bar z\,  y^i y^j}{(y^2+z\bar z)^5}
\end{equation}
Comparing this expression with equation \eqref{TDcorr} we notice the absence of a term proportional to $\d^{ij}$ which implies that, regardless of the explicit form of the displacement operator, the most singular part of the defect OPE of $T^{ij}$ has to vanish. The immediate consequence of that is 
\begin{equation}\label{conjscalar}
 b^1_{DT}-b^2_{DT}=0 \qquad \Rightarrow \qquad  C_D(n)=\frac{24\, h_n}{\pi}
\end{equation}
for any value of $n$. Hence we have verified that \eqref{eq:CDhrel} holds for the four-dimensional free scalar!

The result \eqref{eq:TijO} provides also an additional constraint. Indeed comparing with \eqref{TDcorr} we can extract
\begin{equation}
 b^1_{DT}=\sum_{k=0}^n \frac{c_k k (n-k)}{2n^4 \pi^3}
\end{equation}
and using \eqref{b2}, \eqref{b2b3cd}, \eqref{conjscalar} and  the value of $h_n$ for a free scalar in four dimensions \cite{Hung:2014npa}
 \begin{equation}
 h_n=\frac{n^4-1}{720\, \pi\, n^3},
\end{equation}
we obtain
\begin{equation}
 b^1_{DT}=\frac{n^4-1}{60 n^4 \pi^3}
\end{equation}
Equating the two expressions for $b^1_{DT}$ we get
\begin{equation}
 \sum_{k=0}^n c_k k (n-k)=\frac{n(n^4-1)}{30}\, .
\end{equation}
Once more, rather non-trivially, the r.h.s. is an integer.
Since we have determined the exact value of $C_D$ we can also use the two-point function of the displacement to put a quadratic constraint on the coefficients
\begin{equation}
 \sum_{k=0}^n c_k^2=\frac{n(n^4-1)}{30}
\end{equation}

The constraints collected so far allow to compute the exact values of $c_k$ for $n\leq 5$. The result is
\begin{align}
 n&=2 & c_0&=c_2=0 \qquad c_1=1\\
 n&=3 & c_0&=c_3=0 \qquad c_1=c_2=2\\
 n&=4 & c_0&=c_4=0 \qquad c_1=c_3=3\qquad c_2=4\\
 n&=5 & c_0&=c_5=0 \qquad c_1=c_4=4\qquad c_2=c_3=6
\end{align}
Based on these results and on the structure of the constraints it is very natural to assume that the coefficient $c_k$ are integers. In this case they admit the unique solution
\begin{align}
 c_k=k(n-k) \qquad \Rightarrow \qquad D=\frac{1}{2\pi n^2}\sum_{k=0}^n k(n-k) O_{(\frac{k}{n},\frac{n-k}{n})}
\end{align}
It may be possible to explicitly verify this expression for higher values of $n\,(>5)$ by examining the two-point function of the displacement with the higher spin currents \cite{future}. 

\vfill
\newpage

\bibliographystyle{JHEP-2}

\bibliography{bibliography}

\end{document}